\documentclass[aps,onecolumn,pra,showpacs,amsmath,amssymb,showkeys,10pt]{revtex4-2}
\usepackage{graphicx}
\usepackage{epstopdf}
\usepackage{diagbox}
\pdfoutput=1
\usepackage{braket}
\usepackage{hyperref}
\usepackage{longtable}
\usepackage[ruled,vlined]{algorithm2e}

\begin{document}

	\title{Non-monotonic population scaling and re-entrant photon number dynamics in the dissipative-pumped Tavis--Cummings model}
	
	\author{Hui-hui Miao}
	\email[Correspondence to: Vorobyovy Gory 1, Moscow, 119991, Russia. E-mail address: ]{hhmiao@cs.msu.ru}
	\affiliation{Faculty of Computational Mathematics and Cybernetics, Lomonosov Moscow State University, Vorobyovy Gory 1, Moscow, 119991, Russia}

	\date{\today}

	\begin{abstract}
	This paper investigates nonlinear dynamics in the dissipative-pumped Tavis--Cummings model (TCM). Under global excitation constraints, the competition between photon decay and pumping leads to two significant effects. First, the steady-state peak population exhibits a non-monotonic plateau: contrary to the expected monotonic decay trend at high energy levels, intermediate energy levels show comparable peak populations, and this plateau gradually shifts to higher energies and broadens as the number of atoms increases. Second, the average free photon number exhibits re-entrant dynamics. Under strong pumping, the photon number experiences an initial rise and decay, then slowly increases again, exceeding the initial peak when the system is sufficiently large, with clear thresholds in both system size and pumping rate. To reveal these phenomena, we develop a distributed computing framework suitable for large Hilbert spaces to solve the Lindblad master equation. By exploiting the sparsity of the jump operator and combining it with Cannon's algorithm, we reduce the complexity of the non-unitary terms from $\mathcal{O}(MN^3)$ to $\mathcal{O}(MN)$. Furthermore, the dynamic subspace construction method effectively eliminates redundant quantum states and significantly compresses the Hilbert space. Although the scalability of unitary evolution is limited by communication overhead, the framework as a whole still achieves efficient parameter space exploration. This study demonstrates that the dissipative-pumped TCM serves as a rich platform for studying non-equilibrium nonlinear dynamics and provides a practical numerical tool for its investigation.
	\end{abstract}


	\maketitle

	\section{Introduction}
	\label{sec:Intro}

	Modeling high-dimensional quantum systems represents a central challenge in contemporary computational mathematics, with particular relevance to polymer chemistry and macromolecular biology~\cite{McArdle2020, Baiardi2023, Albuquerque2021}. In simulations of chemical and biological processes, a large number of particles are typically involved. The dimension of the corresponding quantum system grows exponentially with the particle count, a phenomenon widely recognized as the curse of dimensionality~\cite{Bellman1957, Bellman1961}. This issue remains a fundamental obstacle in the study of high-dimensional quantum systems. Recent advances in supercomputing have enabled distributed architectures to partially address the memory and efficiency limitations imposed by high dimensionality. However, beyond these computational challenges, high-dimensional quantum systems also exhibit rich nonlinear dynamical phenomena that remain to be fully explored.

	A key contribution of this work lies in its focus on cavity quantum electrodynamics (QED) models~\cite{Rabi1936, Rabi1937, Dicke1954, Hopfield1958, Casanova2010, Haroche2013, Gu2017, Kockum2019, KockumMiranowicz2019, Forn-Diaz2019}. These models are experimentally feasible and offer a well-established framework for investigating light--matter interactions. Within this paradigm, impurity two-level systems---commonly referred to as atoms---are coupled to a cavity field. Representative models include the Jaynes--Cummings model (JCM)~\cite{Jaynes1963}, the Tavis--Cummings model (TCM)~\cite{Tavis1968}, and their various extensions~\cite{Angelakis2007}. Many recent studies have advanced the field of QED models, exploring phase transitions~\cite{Prasad2018, Wei2021}, quantum many-body phenomena~\cite{OzhigovYI2020, Smith2021}, quantum gate implementations~\cite{Dull2021}, and quantum correlations~\cite{Miao2024, MiaoLi2025}. In our previous work, several strategies were developed to overcome computational difficulties arising from the curse of dimensionality~\cite{You2024, MiaoOzhigov2024, LiMiao2024}. The present study extends this line of research by focusing on the distributed optimization of the Lindblad master equation under the Markov approximation.

	Beyond the computational challenges, the dissipative-pumped Tavis--Cummings model exhibits rich nonlinear dynamical phenomena. The competition between decay (photon leakage) and pumping (photon injection) drives the system far from equilibrium, giving rise to a variety of emergent behaviors that have attracted considerable attention in recent years. For instance, the Jaynes--Cummings model has been shown to exhibit strong nonlinearities in its energy level structure, which have been experimentally verified in circuit QED systems~\cite{Fink2008}. The interplay between dissipation and driving can lead to bistability and dynamical phase transitions in cavity QED systems~\cite{Dombi2015}. More generally, the Tavis--Cummings model and its descendants represent a paradigmatic platform for exploring nonlinear light--matter interactions, including phenomena such as collapses and revivals, entanglement generation, and non-equilibrium steady states~\cite{Larson2021}. The Dicke model (closely related to the Tavis--Cummings model) exhibits the critical properties of the superradiant phase transition and the distinction between equilibrium and non-equilibrium conditions~\cite{Kirton2019}. In our model, the competition between decay and pumping leads to non-equilibrium steady states, population cascades, and the striking re-entrant dynamics we report below. The steady-state peak population exhibits a non-monotonic plateau at intermediate energy levels, while the mean photon number displays a counterintuitive secondary rise after an initial decay. Understanding these nonlinear effects requires systematic parameter-space exploration, which in turn demands efficient numerical methods capable of handling large Hilbert spaces. In this work, we investigate these nonlinear phenomena while developing a distributed computing framework to enable large-scale simulations.
	
	The remainder of this paper is organized as follows. Section~\ref{sec:Model} introduces the target TCM and describes the dynamic subspace construction method. Section~\ref{sec:Theory} presents the theoretical framework, including the Lindblad master equation and its decomposition into unitary and non-unitary terms. Section~\ref{sec:Results} presents numerical results on dissipative-pumped dynamics, including non-monotonic plateau formation, re-entrant photon number dynamics, and efficiency analysis. Finally, Section~\ref{sec:Conclusion} summarizes our findings and discusses future directions. Additional details on the numerical methods, algorithm implementations, and supplementary results are provided in Appendices~\ref{appxsec:DynamicSubspace}--\ref{appxsec:Breakdown}.
	
	\section{Target model}
	\label{sec:Model}
	
	\subsection{Tavis--Cummings model with two-level atoms}
	\label{subsec:Model}
	
	\begin{figure}
		\centering
		\includegraphics[width=.6\textwidth]{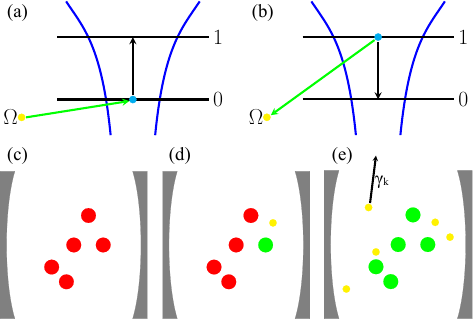} 
		\caption{(online color) {\it TCM with two-level atoms.} Panel (a) shows the excitation process, and panel (b) shows the de-excitation process. The initial state is shown in panel (c), where there exist $N_\mathrm{atoms}$ (for example, $N_\mathrm{atoms}=5$) excited atoms and no photons. In panel (d), an excited atom de-excites and becomes a ground-state atom, at which point a photon is released. In panel (e), when all atoms have transitioned to the ground state, there are $N_\mathrm{atoms}$ photons in the optical cavity. As soon as a photon is released, it can escape into the external environment through a dissipation channel. In the figure, blue circles represent electrons, yellow circles represent photons, red circles represent excited atoms, and green circles represent ground-state atoms. (Adapted from our previous work~\cite{MiaoOzhigov2024}.)}
		\label{fig:TCModel}
	\end{figure}
	
	In this paper, we introduce the TCM involving a large number of two-level atoms. This model was studied in our earlier work~\cite{MiaoOzhigov2024} on unitary evolution, and we now extend our analysis to its non-unitary evolution. The interaction between the atom and the field is explained in detail in Fig.~\ref{fig:TCModel}. In this work, we investigate whether photon pumping can generate nonlinear phenomena. To avoid the influence of photonic coherence effects on the results, we assume that the photons released by different atoms upon de-excitation are non-identical. The study of the combined effects of coherent photon pumping and coherence on nonlinear phenomena will be addressed in future work. The basic states are as follows
	\begin{equation}
		\label{eq:BasisTCModel}
		|\Psi\rangle=\bigotimes_{i=1}^{N_\mathrm{atoms}}|p_i\rangle_{\text{ph}_i}|l_i\rangle_{\text{at}_i},
	\end{equation}
	where $N_\mathrm{atoms}$ is the number of atoms. For the $i$-th atom, $p_i\in\{0,1\}$ denotes the number of free photons, and $l_i\in\{0,1\}$ denotes the electronic state: $l_i=0$ indicates that the electron is in the ground state, and $l_i=1$ indicates that it is in the excited state.
	
	We construct the Hamiltonian under the rotating wave approximation (RWA)~\cite{Wu2007}:
	\begin{equation}
		\label{eq:Hamiltonian}
		\hat{H}_{\text{TCM}}^{\text{RWA}}=\sum_{i=1}^{N_\mathrm{atoms}}\left[\hbar\Omega_i\hat{a}_i^{\dag}\hat{a}_i+\hbar\Omega_i\hat{\sigma}_i^{\dag}\hat{\sigma}_i+g_{\Omega_i}\left(\hat{a}_i^{\dag}\hat{\sigma}_i+\hat{a}_i\hat{\sigma}_i^{\dag}\right)\right],
	\end{equation}
	where $\hbar$ is the reduced Planck constant, $\Omega_i$ is the photon mode, and $g_{\Omega_i}$ is the coupling strength between the field and the electron in the atom. Here, $\hat{a}_i$ is the photon annihilation operator, $\hat{a}_i^{\dag}$ is the photon creation operator, $\hat{\sigma}_i$ is the electron relaxation operator, and $\hat{\sigma}_i^{\dag}$ is the electron excitation operator. We set $\Omega_1=\Omega_2=\cdots=\Omega_{N_\text{atoms}}=\Omega$ and $g_{\Omega_1}=g_{\Omega_2}=\cdots=g_{\Omega_{N_\text{atoms}}}=g$.
	
	\subsection{Method for constructing the dynamic subspace}
	\label{subsec:DynamicSubspace}
	
	 Theoretically, for a single-atom system, the Hamiltonian constructed using the traditional tensor product method has dimension 4 (because the Hilbert space has four quantum states: $|00\rangle$, $|01\rangle$, $|10\rangle$, and $|11\rangle$). In practice, we can reduce the dimension by constructing a dynamical subspace. In our earlier work, we referred to this as the "generator algorithm"~\cite{Miao2023}. The core idea of this method is to derive all possible quantum states based on the constraints and initial conditions of the model itself. Consider, for example, a single-atom system. If the initial state is $|01\rangle$, then through interactions and dissipation, we can obtain $|00\rangle$ and $|10\rangle$, while $|11\rangle$ is never populated. Ultimately, the reduced Hilbert space contains only three states, and the Hamiltonian dimension is reduced from 4 to 3. The more atoms there are, the more pronounced the advantage of the dynamical subspace construction method becomes. For example, when $N_\mathrm{atoms}=10$, the dimension of the reduced Hamiltonian is only $5.63\%$ of the full Hamiltonian dimension, while the memory usage is only $0.32\%$ of that for the full Hamiltonian. For details on the dynamical subspace construction method and its efficiency analysis for a system of $N_\mathrm{atoms}$ two-level atoms, see Appendix~\ref{appxsec:DynamicSubspace}.
	
	\section{Numerical method and distributed implementation}
	\label{sec:Theory}
	
	\subsection{Lindblad master equation}
	\label{subsec:QME}
	
	Under the Markovian approximation, the quantum master equation (QME) is widely used to describe the dissipative-pumped dynamics of cavity QED models. The QME is a standard tool for studying open quantum systems~\cite{Breuer2002} and is consistent with the principles of quantum thermodynamics~\cite{Alicki1979, Kosloff2013}. The QME for the density matrix $\hat{\rho}$ of the system takes the form:
	\begin{equation}
		\label{eq:QME}
		i\hbar\dot{\hat{\rho}}=\hat{\mathcal{L}}\left(\hat{\rho}\right)=\left[\hat{H},\hat{\rho}\right]+i\hat{L}\left(\hat{\rho}\right),
	\end{equation}
	where $\hat{H}$ is the system Hamiltonian, $\hat{\mathcal{L}}\left(\hat{\rho}\right)$ is the Liouvillian superoperator, $\hat{L}\left(\hat{\rho}\right)$ is the Lindblad superoperator, and $\left[\hat{H},\hat{\rho}\right]=\hat{H}\hat{\rho}-\hat{\rho}\hat{H}$ denotes the commutator. The possible dissipative transitions are represented by a graph $\mathcal{K}$. The vertices of this graph correspond to the system states, and the edges indicate the allowed dissipative channels. Each dissipation channel has a corresponding inverse (pumping) channel. $\hat{L}\left(\hat{\rho}\right)$ is given by
	\begin{equation}
		\label{eq:LindbladOperator}
		\hat{L}\left(\hat{\rho}\right)=\sum_{k\in \mathcal{K}}\left(\hat{L}_k\left(\hat{\rho}\right)+\hat{L'}_k\left(\hat{\rho}\right)\right).
	\end{equation}
	Here, $\hat{L}_k\left(\hat{\rho}\right)$ is the dissipation superoperator acting on the density matrix $\hat{\rho}$, associated with the jump operator $\hat{A}_k$:
	\begin{equation}
		\label{eq:DissSuper}
		\hat{L}_k\left(\hat{\rho}\right)=\gamma_k\left(\hat{A}_k\hat{\rho}\hat{A}_k^\dagger-\frac{1}{2}\left\{\hat{\rho},\hat{A}_k^\dagger\hat{A}_k\right\}\right),
	\end{equation}
	where $\gamma_k$ (with $k\in\mathcal{K}$) is the total rate of spontaneous photon emission due to photon loss from the cavity to the external environment (dissipative rate), and $\left\{\hat{\rho},\hat{A}_k^\dagger\hat{A}_k\right\}=\hat{\rho}\hat{A}_k^\dagger\hat{A}_k+\hat{A}_k^\dagger\hat{A}_k\hat{\rho}$ denotes the anticommutator. Correspondingly, $\hat{L'}_k\left(\hat{\rho}\right)$ is the pumping superoperator, given by:
	\begin{equation}
		\label{eq:PumpSuper}
		\begin{aligned}
			\hat{L'}_k\left(\hat{\rho}\right)&=\gamma_k'\left(\hat{A}_k^\dagger\hat{\rho}\hat{A}_k-\frac{1}{2}\left\{\hat{\rho},\hat{A}_k\hat{A}_k^\dagger\right\}\right)\\
			&=\gamma_k'\left(\hat{A'}_k\hat{\rho}\hat{A'}_k^\dagger-\frac{1}{2}\left\{\hat{\rho},\hat{A'}_k^\dagger\hat{A'}_k\right\}\right),
		\end{aligned}
	\end{equation}
	where $\gamma_k'$ is the total spontaneous photon influx rate (pumping rate), with $k\in\mathcal{K}$. In Eq.~\eqref{eq:PumpSuper}, we use $\hat{A'}_k$ instead of $\hat{A}_k^\dagger$. This choice maintains consistency with the structure of Eq.~\eqref{eq:DissSuper} and facilitates the separate discussion of dissipative and pumping channels.
	
	The solution $\hat{\rho}\left(t\right)$ to Eq.~\eqref{eq:QME} is obtained approximately via a two-step procedure. First, the unitary term of Eq.~\eqref{eq:QME} is propagated forward by a single time step:
	\begin{equation}
		\label{eq:UnitaryTerm}
		\tilde{\hat{\rho}}\left(t+\Delta t\right)=\exp\!\left({-\frac{i}{\hbar}\hat{H}\Delta t}\right)\hat{\rho}\left(t\right)\exp\!\left(\frac{i}{\hbar}\hat{H}\Delta t\right),
	\end{equation}
where $\Delta t$ is the time step. Second, the non-unitary term of Eq.~\eqref{eq:QME} is propagated forward by one step:
	\begin{equation}
		\label{eq:Non-unitaryTerm}
		\hat{\rho}\left(t+\Delta t\right)=\tilde{\hat{\rho}}\left(t+\Delta t\right)+\frac{1}{\hbar}\hat{L}\left(\tilde{\hat{\rho}}(t+\Delta t)\right)\Delta t.
	\end{equation}
	
	The dimensions of the Hamiltonian and the density matrix grow exponentially with the number of atoms, which makes it difficult to mathematically model large-dimensional quantum systems using the QME. This core challenge in quantum many-particle physics is known as the curse of dimensionality. Fortunately, recent advances in supercomputing have enabled us to mitigate this problem. The key is to develop algorithms for solving the QME on supercomputing platforms. Therefore, we need to perform distributed computing separately for the unitary contribution, $\tilde{\hat{\rho}}(t+\Delta t)$ in Eq.~\eqref{eq:UnitaryTerm}, and the non-unitary contribution, $\frac{1}{\hbar}\hat{L}\left(\tilde{\hat{\rho}}(t+\Delta t)\right)\Delta t$ in Eq.~\eqref{eq:Non-unitaryTerm}.
	
	\subsection{Unitary term}
	\label{subsec:Unitary}
	
	To overcome the memory bottleneck caused by high-dimensional large matrices, we employ the Cannon's algorithm~\cite{Cannon1969} for core matrix multiplication and addition operations. Our previous research~\cite{MiaoOzhigov2024, LiMiao2024} have previously applied this algorithm to solve the unitary terms of the QME. The core idea of the algorithm is to divide a large matrix into several sub-blocks and distribute them across multiple processors. By having each processor receive and compute only its corresponding sub-block, the algorithm significantly reduces the memory requirements for each individual processor. As the number of processors increases, the communication overhead between processors---and even between nodes---increases significantly, becoming a key factor that limits the algorithm's scalability. In order to apply a distributed computing algorithm, the matrix exponential parts of Eq.~\eqref{eq:UnitaryTerm} need to be converted into matrix multiplication and addition operations. In the past few decades, many calculation methods for exponential matrices have been proposed~\cite{Sidje1998, Moler2003}. We adopt the Taylor series approximation:
	\begin{equation}
		\label{eq:Taylor}
		\exp(A)=\sum_{k=0}^{\infty}\frac{A^k}{k!},
	\end{equation}
	where $A$ is a matrix. Now Eq.~\eqref{eq:UnitaryTerm} can be rewritten as follows:
	\begin{equation}
		\label{eq:UnitaryPartTaylor}
		\tilde{\hat{\rho}}\left(t+\Delta t\right)=\left[\sum_{k=0}^{\infty}\frac{(-\frac{i}{\hbar}\hat{H}\Delta t)^k}{k!}\right]\hat{\rho}\left(t\right)\left[\sum_{k=0}^{\infty}\frac{(\frac{i}{\hbar}\hat{H}\Delta t)^k}{k!}\right].
	\end{equation}
	In practice, the infinite series is truncated at a finite order. We have found that a truncation order of $k_{\text{max}}=10$ is sufficient to ensure numerical accuracy.
	
	\begin{figure}
		\centering
        \includegraphics[width=.35\textwidth]{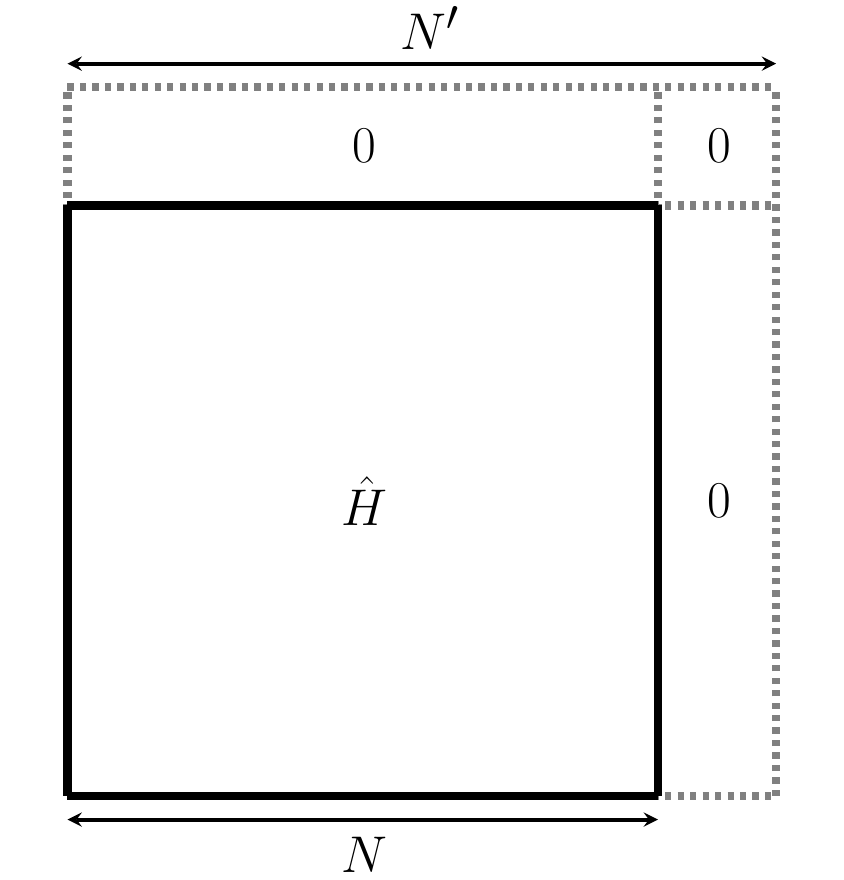}
        \caption{(online color) {\it Hamiltonian expansion.} The original Hamiltonian, represented as a black square, is padded with zeros (gray rectangles) to obtain an expanded Hamiltonian of dimension $N'$, where $N'$ is chosen to be divisible by the processor grid dimension $p_x$. The same padding procedure is applied to the density matrix.}
        \label{fig:HamiltonianExpansion}
	\end{figure}
	
	\begin{figure}
		\centering
		\includegraphics[width=1.\textwidth]{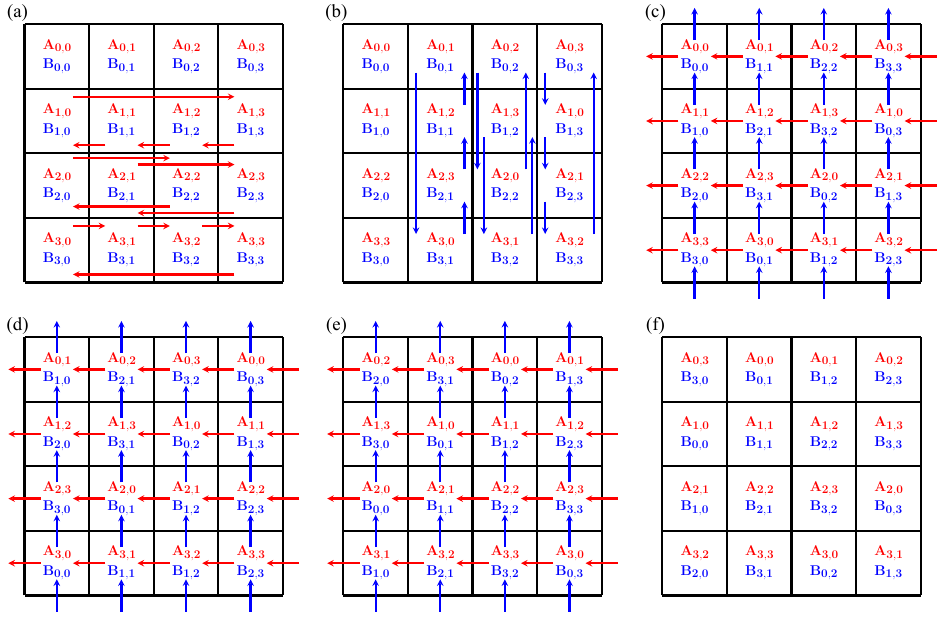} 
		\caption{(online color) {\it Illustration of Cannon's algorithm on a $4\times4$ processor grid.} Panel (a): Initial distribution of blocks $A_{i,j}$ and $B_{i,j}$ on each processor $P_{i,j}$, followed by horizontal alignment (shifting $A_{i,j}$ left by $i$ steps). Panel (b): Vertical alignment: shifting $B_{i,j}$ up by $j$ steps. Panels (c)--(e): Three iterations of cyclic shifts and multiply-accumulate operations: after each multiplication step, $A$ blocks are shifted left cyclically and $B$ blocks are shifted up cyclically. Panel (f): Final block distribution after completing all iterations, yielding the product matrix $C=A\times B$. (Adapted from our previous work~\cite{MiaoOzhigov2024}.)}
		\label{fig:Unitary}
	\end{figure}
	
	Assuming the total number of processors is $p_{\mathrm{total}}$, we arrange them in a two-dimensional grid with $p_x=p_y=\sqrt{p_{\mathrm{total}}}$. Each matrix is then partitioned into $p_x \times p_y$ sub-blocks. If the dimension is not divisible by $p_x$ (and hence by $p_y$), both the Hamiltonian and the density matrix are padded with zeros to a dimension that is divisible (see Fig.~\ref{fig:HamiltonianExpansion}). Initially, the matrix block assigned to processor $P_{i,j}$ is denoted by $A_{i,j}$ (or $B_{i,j}$), where $0\leq i,j<p_x$. The procedure of Cannon's algorithm is outlined as follows:
	\begin{itemize}
		\item Initialization: Each processor $P_{i,j}$ initially holds two submatrices, $A_{i,j}$ and $B_{i,j}$, derived from the full matrices $A$ and $B$, respectively.
		\item Row-wise alignment: Each submatrix $A_{i,j}$ is cyclically shifted left by $i$ positions (see Fig.~\ref{fig:Unitary}(a)).
		\item Column-wise alignment: Each submatrix $B_{i,j}$ is then cyclically shifted upward by $j$ positions (see Fig.~\ref{fig:Unitary}(b)).
		\item Iterative computation and shift: The core iteration begins. Each processor performs a multiply-accumulate operation on its current $A$ and $B$ blocks. Following this, every $A$ block is passed to the left neighboring processor (with wrap-around), and every $B$ block is passed to the upward neighboring processor (with wrap-around).
		\item Final result: The iteration step (multiplication followed by cyclic shifts) is repeated $p_x$ times. After the final multiplication, the product matrix $C = A \times B$ is obtained (see Fig.~\ref{fig:Unitary}(c)--(f)).
	\end{itemize}
	
	Let $\operatorname{Cannon}(A_1, A_2)$ denote Cannon's algorithm for multiplying two square matrices of equal dimension. We define recursively:
	\begin{equation}
		\label{eq:CannonFunc}
		\operatorname{Cannon}^k(A_1, \dots, A_k) = \operatorname{Cannon}(A_1, \operatorname{Cannon}^{k-1}(A_2, \dots, A_k)),
	\end{equation}
	with $\operatorname{Cannon}^1(A)=A$. If all arguments are equal to $A$, we write $\operatorname{Cannon}^k(A)$.
	
	Now we can transform Eq.~\eqref{eq:UnitaryPartTaylor} into a distributed computing form. First, we compute the left part of Eq.~\eqref{eq:UnitaryPartTaylor}:
	\begin{equation}
		\label{eq:TaylorLeft}
		\hat{L}_{\text{exp}}=\hat{I}-\frac{i}{\hbar}\hat{H}\Delta t+\sum_{k=2}^{10}\frac{\operatorname{Cannon}^k(-\frac{i}{\hbar}\hat{H}\Delta t)}{k!},
	\end{equation}
	where $\hat{I}$ is the identity matrix. Second, we compute the right part of Eq.~\eqref{eq:UnitaryPartTaylor}:
	\begin{equation}
		\label{eq:TaylorRight}
		\hat{R}_{\text{exp}}=\hat{I}+\frac{i}{\hbar}\hat{H}\Delta t+\sum_{k=2}^{10}\frac{\operatorname{Cannon}^k(\frac{i}{\hbar}\hat{H}\Delta t)}{k!}.
	\end{equation}
	The distributed computing transformation of the numerical method for quantum unitary evolution is now complete:
	\begin{equation}
		\label{eq:UnitaryPartTaylorCannon}
		\tilde{\hat{\rho}}\left(t+\Delta t\right)=\operatorname{Cannon}^3\bigl(\hat{L}_{\text{exp}},\hat{\rho}\left(t\right),\hat{R}_{\text{exp}}\bigr).
	\end{equation}
	
	\subsection{Non-unitary term}
	\label{subsec:Non-unitary}
	
	\begin{figure}
		\centering
		\includegraphics[width=1.\textwidth]{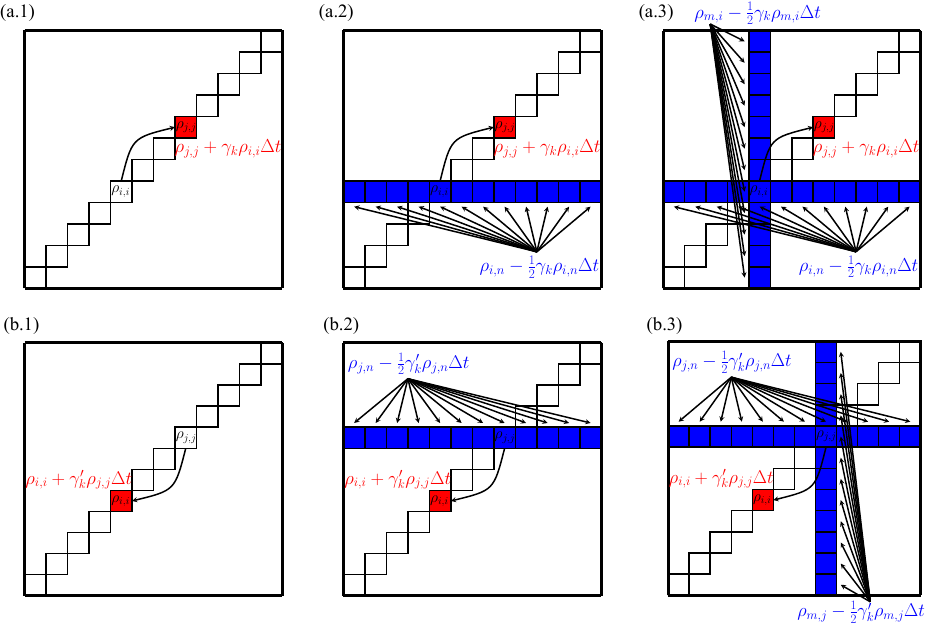} 
		\caption{(online color) {\it Numerical update for the non-unitary term with dissipation channel $\hat{A}_k=|j\rangle\langle i|$ and its reverse process.} Panels (a.1)--(a.3) show the dissipative process, and panels (b.1)--(b.3) show its inverse process. All operations are applied to $\hat{\rho}(t)$ to obtain $\hat{\rho}(t+\Delta t)$. Panel (a.1): Population transfer from the diagonal element of state $|i\rangle$ to that of state $|j\rangle$, proportional to the original population of $|i\rangle$. Panel (a.2): Subtraction applied to all elements in the $i$-th row, each reduced by a factor proportional to itself. Panel (a.3): Subtraction applied to all elements in the $i$-th column, each reduced by a factor proportional to itself. Panel (b.1): Reverse population transfer from the diagonal element of state $|j\rangle$ to that of state $|i\rangle$, proportional to the original population of $|j\rangle$. Panel (b.2): Subtraction applied to all elements in the $j$-th row, each reduced by a factor proportional to itself. Panel (b.3): Subtraction applied to all elements in the $j$-th column, each reduced by a factor proportional to itself.}
		\label{fig:Non-unitary}
	\end{figure}
		
	\begin{figure}
		\centering
		\includegraphics[width=1.\textwidth]{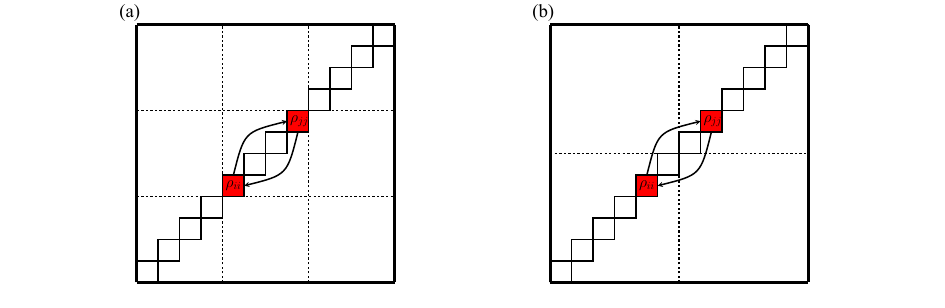} 
		\caption{(online color) {\it The impact of different processor grids on dissipation channels.} Panel (a) is a $3\times3$ processor grid, and panel (b) is a $2\times2$ processor grid.}
		\label{fig:Grids}
	\end{figure}
	
	The detailed steps of the numerical update for the non-unitary term with $\hat{A}_k=|j\rangle\langle i|$ are shown below. All operations are applied to $\tilde{\hat{\rho}}(t+\Delta t)$ to obtain $\hat{\rho}(t+\Delta t)$. In this paper, we set $\hbar=1$. The dissipative process is shown in panels (a.1)--(a.3) of Fig.~\ref{fig:Non-unitary}: 
	\begin{itemize}
		\item Population transfer from $\tilde{\hat{\rho}}_{i,i}(t+\Delta t)$ to $\tilde{\hat{\rho}}_{j,j}(t+\Delta t)$:
		\begin{equation}
			\label{eq:Non-unitary-1}
    			\hat{\rho}_{j,j}(t+\Delta t)=\tilde{\hat{\rho}}_{j,j}(t+\Delta t)+\gamma_k\Delta t\tilde{\hat{\rho}}_{i,i}(t+\Delta t).
    		\end{equation}
		\item Subtraction along the $i$-th row: for all $n$,
		\begin{equation}
			\label{eq:Non-unitary-2}
    			\hat{\rho}_{i,n}(t+\Delta t)=\left(1-\frac{\gamma_k\Delta t}{2}\right)\tilde{\hat{\rho}}_{i,n}(t+\Delta t).
    		\end{equation}
		\item Subtraction along the $i$-th column: for all $m$,
		\begin{equation}
			\label{eq:Non-unitary-3}
	   		\hat{\rho}_{m,i}(t+\Delta t)=\left(1-\frac{\gamma_k\Delta t}{2}\right)\tilde{\hat{\rho}}_{m,i}(t+\Delta t).
   		\end{equation}
   	\end{itemize}
   	The inverse process is shown in panels (b.1)--(b.3) of Fig.~\ref{fig:Non-unitary}:
   	\begin{itemize}
		\item Population transfer from $\tilde{\hat{\rho}}_{j,j}(t+\Delta t)$ to $\tilde{\hat{\rho}}_{i,i}(t+\Delta t)$:
		\begin{equation}
			\label{eq:Non-unitary-4}
    			\hat{\rho}_{i,i}(t+\Delta t)=\tilde{\hat{\rho}}_{i,i}(t+\Delta t)+\gamma_k'\Delta t\tilde{\hat{\rho}}_{j,j}(t+\Delta t).
    		\end{equation}
		\item Subtraction along the $j$-th row: for all $n$,
		\begin{equation}
			\label{eq:Non-unitary-5}
	    		\hat{\rho}_{j,n}(t+\Delta t)=\left(1-\frac{\gamma_k'\Delta t}{2}\right)\tilde{\hat{\rho}}_{j,n}(t+\Delta t).
    		\end{equation}
		\item Subtraction along the $j$-th column: for all $m$,
		\begin{equation}
			\label{eq:Non-unitary-6}
    			\hat{\rho}_{m,j}(t+\Delta t)=\left(1-\frac{\gamma_k'\Delta t}{2}\right)\tilde{\hat{\rho}}_{m,j}(t+\Delta t).
    		\end{equation}
	\end{itemize}
	
	As shown in Eq.~\eqref{eq:QME}, calculating non-unitary terms involves matrix multiplication, with a computational complexity of $\mathcal{O}(MN^3)$, where $M=\dim\left(\mathcal{K}\right)$ is the number of dissipation channels and $N=\dim\left(\hat{H}\right)$ is the dimension of the Hamiltonian. However, as illustrated in Fig.~\ref{fig:Non-unitary}, by leveraging the sparsity of the jump operator, we can simplify the matrix operations to a point operation, a row operation, and a column operation, regardless of whether the channel is dissipation or pumping. Therefore, the overall computational complexity is reduced to $\mathcal{O}(MN)$. For a detailed description of the process that reduces the computational complexity of solving non-unitary terms from $\mathcal{O}(MN^3)$ to $\mathcal{O}(MN)$, see Appendix~\ref{appxsec:Non-unitary}. This method significantly improves computational efficiency. Furthermore, since only the point operation involves data transfer, while the row and column operations use only local data, distributed computation of non-unitary terms becomes easier. This approach avoids the high cross-processor communication costs associated with transferring entire blocks via the Cannon's algorithm, as required for solving unitary terms. Additionally, as shown in Fig.~\ref{fig:Grids}, whether a dissipation channel involves cross-processor communication depends on the processor network being used. This suggests an optimization strategy: the cross-processor communication cost of non-unitary terms can be reduced by selecting an optimal processor network. An even better approach is to minimize this cost by rearranging the quantum states along the diagonal of the Hamiltonian and density matrix through row and column permutations.
	
	\section{Results and discussion}
	\label{sec:Results}
	
	Before presenting the results, we first clarify the concept of energy levels under global excitation constraints. Let $E = \hbar\Omega$ represent the energy of a single excited-state atom (or a single free photon). Initially, we have $N_{\mathrm{atoms}}$ excited atoms and no photons, so the initial total energy is $N_{\mathrm{atoms}}E$. Each photon mode can contain at most one photon (hard-core boson constraint). Pumping can inject excitations into the system, but the total energy cannot exceed the initial energy $N_{\mathrm{atoms}}E$. Therefore, the system energy level $N_\mathrm{exc}E$ corresponds to configurations with $N_\mathrm{exc}$ total excitations (excited atoms plus free photons), where $N_\mathrm{exc}=0,1,\dots,N_{\mathrm{atoms}}$. In the purely dissipative case, $N_\mathrm{exc}$ decreases over time as photons leak out. In the dissipative-pumped case, pumping can increase $N_\mathrm{exc}$ toward higher values, but never beyond $N_{\mathrm{atoms}}$.

	In this work, the initial state consists of $N_{\mathrm{atoms}}$ excited atoms and zero photons. Therefore, the initial total number of excitations is $N_{\mathrm{atoms}}$, and the initial system energy level is $N_{\mathrm{atoms}}E$.
	
	\subsection{Dissipative dynamics}
	\label{subsec:DissDynamics}
	
	\begin{figure}
		\centering
		\includegraphics[width=1.\textwidth]{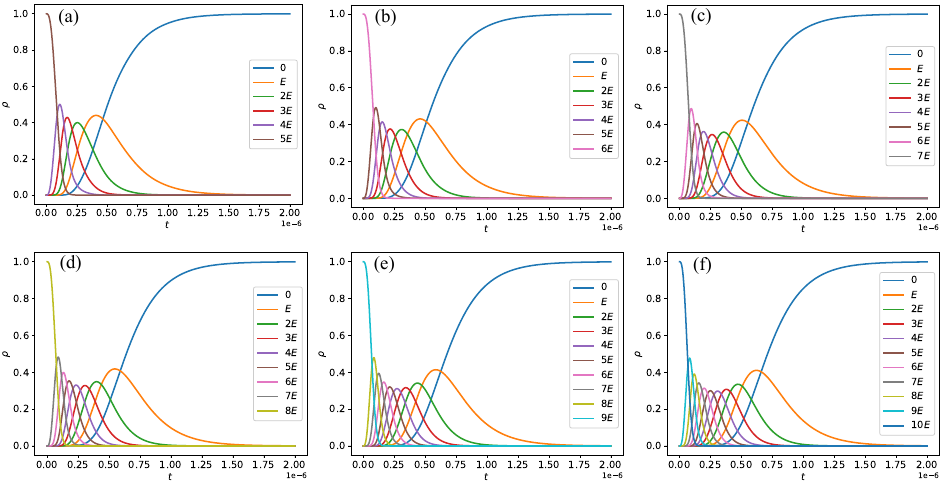} 
		\caption{(online color) {\it Purely dissipative dynamics of the TCM with $N_{\mathrm{atoms}}$ two-level atoms (no pumping).} Panels (a)--(f) correspond to $N_{\mathrm{atoms}} = 5, 6, 7, 8, 9, 10$, respectively.}
		\label{fig:Dissipative}
	\end{figure}

	We first study the purely dissipative dynamics (without pumping) of the target model as a reference. As shown in Fig.~\ref{fig:Dissipative}(a), for an initial state with five excited-state atoms ($N_{\mathrm{atoms}}=5$), the curve for the system energy level $5E$ drops rapidly from 1 to 0. Meanwhile, the curve for $4E$ quickly rises, reaches a peak, and then decays to zero. The curve for $3E$ follows a similar pattern, peaking and then decaying, and this behavior propagates through successively lower energy levels. Eventually, the curve for $0E$ (the ground state) rises and approaches 1, indicating that the system has reached a state with no photons. Similar behavior is observed for $N_{\mathrm{atoms}} = 6, 7, 8, 9, 10$ (Figs.~\ref{fig:Dissipative}(b)--(f)).
	
	This cascade of population from higher to lower energy levels reflects the sequential decay of excited-state atoms: when an atom decays from the excited state to the ground state, it releases a photon, which subsequently leaves the cavity. The energy is thus transferred stepwise down the energy ladder until all atoms are in the ground state and no photons remain. This purely dissipative dynamics serves as a baseline for comparison with the dissipative-pumped case discussed below, where the introduction of pumping fundamentally alters the steady-state behavior.
	
	\subsection{Dissipative-pumped dynamics}
	\label{subsec:DissPumpDynamics}
	
	We now introduce pumping to the system. For convenience, we use $\mu$ (rather than $\gamma_k'$) to denote the pumping rate. The pumping rate is characterized by $0 \leq \mu = \gamma'/\gamma < 1$, where $\gamma$ is the decay rate and $\gamma'$ is the pumping rate. When $\mu>0$, the system is driven away from equilibrium by the competition between photon loss (decay) and external pumping.

\begin{figure}
		\centering
		\includegraphics[width=1.\textwidth]{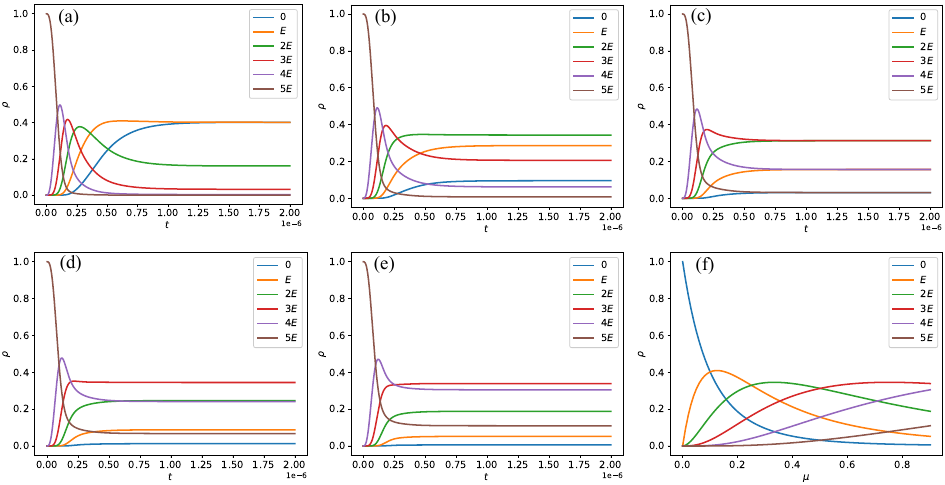} 
		\caption{(online color) {\it Dissipative-pumped dynamics of the TCM with $N_\mathrm{atoms}=5$ two-level atoms.} Panels (a)--(e): Time evolution of energy-level populations for pumping rates $\mu=0.1, 0.3, 0.5, 0.7, 0.9$, respectively.  Due to the competition between decay and pumping, the system reaches a non-equilibrium steady state where multiple energy levels remain populated. Panel (f): Steady-state population distribution as a function of $\mu$ for each energy level ($0$ to $5E$). The cascade of population to higher energy levels becomes more pronounced as $\mu$ increases.}
		\label{fig:Dissipative+Pumped-5Atoms}
	\end{figure}

	Fig.~\ref{fig:Dissipative+Pumped-5Atoms} shows the dissipative-pumped dynamics for the smallest system studied ($N_{\mathrm{atoms}}=5$). Panels (a)--(e) display the time evolution of energy-level populations for pumping rates $\mu = 0.1, 0.3, 0.5, 0.7, 0.9$, respectively. Several key features are observed:
	\begin{itemize}
		\item Non-equilibrium steady state: Unlike the purely dissipative case (see Fig.~\ref{fig:Dissipative}), where all population eventually decays to the ground state, the introduction of pumping leads to a steady state where multiple energy levels remain populated. This reflects the balance between decay and pumping.
		\item Population cascade: As $\mu$ increases, population is progressively pushed to higher energy levels. For $\mu = 0.9$, the highest population appears at $3E$, followed by $4E$, $2E$, and $5E$ (in descending order). As $\mu$ increases from $0.1$ to $0.9$, the population of the $5E$ level (the maximum energy level for $N_{\mathrm{atoms}}=5$) grows monotonically, indicating that stronger pumping drives population to the highest accessible energy levels.
		\item Non-monotonic $\mu$-dependence: Each energy level exhibits a peak in its steady-state population at a characteristic $\mu$ value, reflecting the competition between decay and pumping.
	\end{itemize}

	Panel (f) of Fig.~\ref{fig:Dissipative+Pumped-5Atoms} summarizes the steady-state population distribution as a function of $\mu$ for each energy level from $0$ to $5E$. This panel clearly shows the cascade: as $\mu$ increases, population is transferred from lower to higher energy levels, with each level rising and then falling as the next level becomes dominant.

\begin{figure}
		\centering
		\includegraphics[width=1.\textwidth]{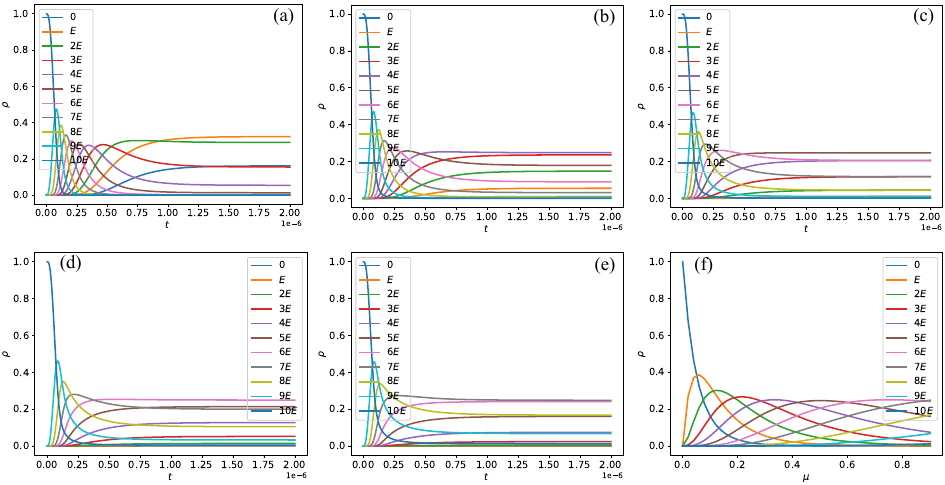} 
		\caption{(online color) {\it Dissipative-pumped dynamics of the TCM with $N_\mathrm{atoms}=10$ two-level atoms.} Same format as Fig.~\ref{fig:Dissipative+Pumped-5Atoms} for $N_\mathrm{atoms}=10$. Panels (a)--(e): Time evolution for $\mu=0.1, 0.3, 0.5, 0.7, 0.9$. Panel (f): Steady-state population distribution versus $\mu$ for energy levels $0$ to $10E$.}
		\label{fig:Dissipative+Pumped-10Atoms}
	\end{figure}

	Fig.~\ref{fig:Dissipative+Pumped-10Atoms} shows the same dynamics for the largest system studied ($N_{\mathrm{atoms}}=10$). The same qualitative features are observed, but with an important difference: the population cascade extends to higher energy levels (up to $10E$).

	The corresponding results for $N_{\mathrm{atoms}}=6,7,8,9$ are provided in Appendix~\ref{appxsec:DissPumpDynamics} for completeness.
	
	\subsection{Non-monotonic steady-state peak population and scaling}
	\label{subsec:NonlinearScaling}

	\begin{figure}
		\centering
		\includegraphics[width=1.\textwidth]{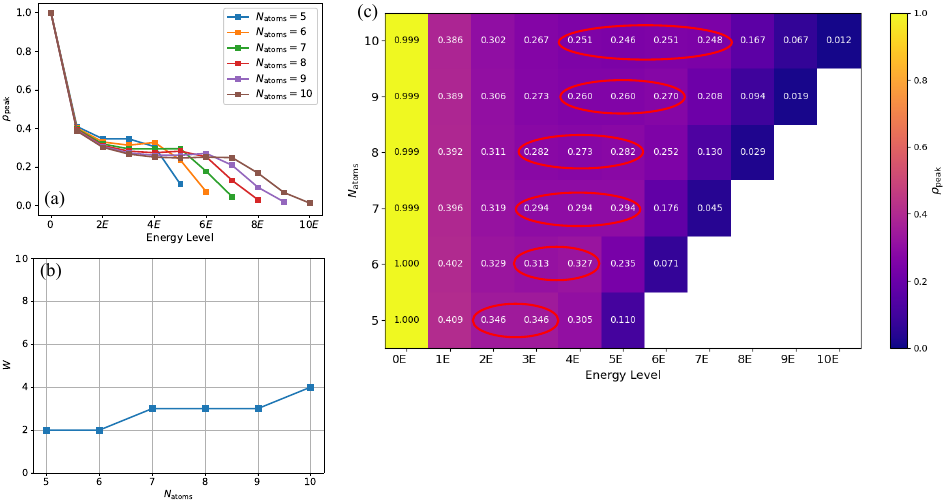} 
		\caption{(online color) {\it Non-monotonic steady-state peak population and its scaling in the dissipative-pumped TCM.} Panel (a): One-dimensional curves of the steady-state peak population $\rho_{\mathrm{peak}}$ as a function of the energy level for $N_{\mathrm{atoms}}=5$ to $10$ (six curves). A non-monotonic plateau emerges at intermediate energy levels for each system size. Panel (b): The plateau width $W$ as a function of $N_\mathrm{atoms}$, demonstrating a clear monotonic increase with system size. Panel (c): Two-dimensional phase diagram (heatmap) of $\rho_\mathrm{peak}$, showing that the plateau shifts to higher energy levels and broadens as $N_\mathrm{atoms}$ increases.}
		\label{fig:PeakPopulation}
	\end{figure}

	Combining the steady-state population distributions from Fig.~\ref{fig:Dissipative+Pumped-5Atoms}(f), Fig.~\ref{fig:Dissipative+Pumped-10Atoms}(f), and the corresponding panels (f) in Appendix~\ref{appxsec:DissPumpDynamics} (Figs.~\ref{appxfig:Dissipative+Pumped-6Atoms}--\ref{appxfig:Dissipative+Pumped-9Atoms}), we now analyze the steady-state peak population $\rho_{\mathrm{peak}}$ for each energy level, defined as the maximum population attained by that energy level as the pumping rate $\mu$ varies from $0$ to $0.9$. Fig.~\ref{fig:PeakPopulation} summarizes this behavior across system sizes from $N_{\mathrm{atoms}}=5$ to $10$.

	\subsubsection{Non-monotonic plateau}
	\label{subsubsec:Plateau}

	Panel (a) of Fig.~\ref{fig:PeakPopulation} shows $\rho_{\mathrm{peak}}$ as a function of energy level for $N_{\mathrm{atoms}}=5$ to $10$ (six curves). Several striking features are observed:
	\begin{itemize}
		\item Ground state dominance at $\mu=0$: Only the ground state (energy level $0$) can reach a population of $1$ (when $\mu=0$), while all other energy levels have zero population. When $\mu>0$, no energy level can achieve a population of $1$.
		\item Optimal pumping rate for each level: Each energy level has a characteristic pumping rate $\mu$ that maximizes its population. In theory, higher energy levels should have lower achievable peak populations.
		\item Non-monotonic plateau: Contrary to the theoretical expectation of monotonic decay, panel (a) reveals a plateau (or even a weak increase) at intermediate energy levels. For $N_{\mathrm{atoms}}=5$, this plateau occurs between $2E$ and $3E$; for larger systems, the plateau shifts to higher energy levels. Beyond the plateau, $\rho_{\mathrm{peak}}$ decays rapidly again at the highest energy levels.
	\end{itemize}

	This non-monotonic behavior challenges the naive expectation that peak population decreases monotonically with energy level. The plateau indicates a range of intermediate energy levels where multiple levels achieve comparable maximum populations under their respective optimal pumping rates, despite the theoretical trend that higher levels should have lower peaks. This anomalous behavior is a signature of the complex competition between decay and pumping in the dissipative-pumped Tavis--Cummings model.

	\subsubsection{Size-dependent scaling}
	\label{subsubsec:SizeScaling}

	Panel (b) of Fig.~\ref{fig:PeakPopulation} shows the plateau width $W$ as a function of $N_{\mathrm{atoms}}$. The plateau width increases monotonically with system size, from $W = 2$ for $N_{\mathrm{atoms}} = 5$ to $W = 4$ for $N_{\mathrm{atoms}} = 10$. This indicates that larger systems support a broader range of energy levels that achieve comparable peak populations.

	Panel (c) of Fig.~\ref{fig:PeakPopulation} presents a two-dimensional phase diagram (heatmap) of $\rho_{\mathrm{peak}}$ with $N_{\mathrm{atoms}}$ on the vertical axis and energy level on the horizontal axis. The color scale represents $\rho_{\mathrm{peak}}$. A clear trend is visible: the plateau region (indicated by the circle) shifts to higher energy levels as $N_{\mathrm{atoms}}$ increases, and its width grows accordingly.

	The observed scaling behavior can be explained as follows. In the dissipative-pumped Tavis--Cummings model, each energy level has an optimal pumping rate that maximizes its population. Theoretically, one might expect higher energy levels to have lower peak populations. However, panel (a) of Fig.~\ref{fig:PeakPopulation} reveals a non-monotonic feature: the peak populations do not decay monotonically with energy level. Instead, a plateau (or weak increase) emerges at intermediate energy levels, where multiple levels achieve comparable maximum populations. As $N_{\mathrm{atoms}}$ increases, this plateau shifts to higher energy levels and broadens. This non-monotonic plateau is a key nonlinear characteristic of the dissipative-pumped Tavis--Cummings model, distinguishing it from simple monotonic decay behavior.
	
	\subsection{Mean free photon number dynamics and steady-state scaling}
	\label{subsec:PhotonNumberDynamics}
	
	\begin{figure}
		\centering
		\includegraphics[width=1.\textwidth]{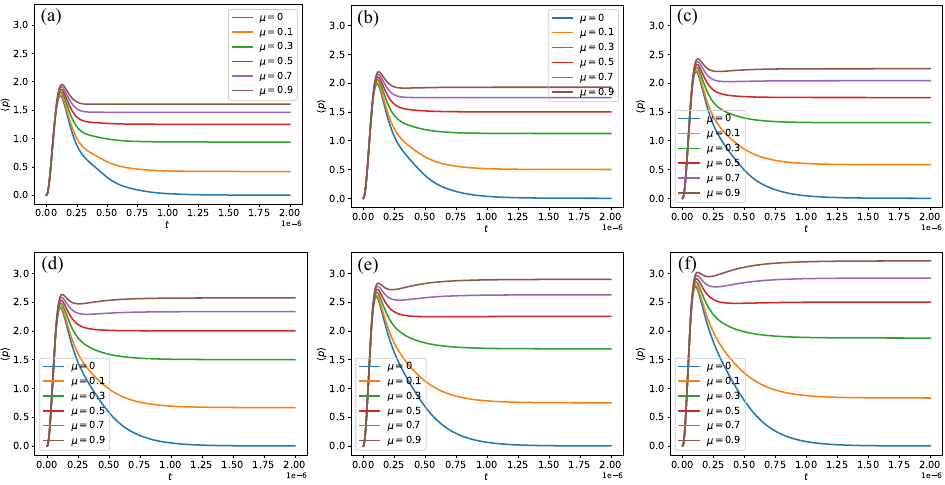} 
		\caption{(online color) {\it Mean free photon number dynamics in the dissipative-pumped TCM.} Panels (a)--(f) correspond to $N_\mathrm{atoms}=5$ to $10$. Curves show pumping rates $\mu=0, 0.1, 0.3, 0.5, 0.7, 0.9$. For small systems ($N_\mathrm{atoms}\leq 6$) or weak pumping, each curve peaks and then decays to a steady state. For larger systems ($N_\mathrm{atoms}\geq 7$) with strong pumping ($\mu = 0.9$), an evident re-entrant behavior emerges: after an initial decay, $\langle p\rangle$ slowly increases again and exceeds the initial peak for $N_\mathrm{atoms}\geq 9$. Similarly, for $\mu=0.7$ when $N_\mathrm{atoms}\geq 8$, an evident re-entrant behavior also emerges. This phenomenon arises from pump-driven photon injection and the subsequent atom-photon interactions.}
		\label{fig:AveragePhotonNumber-T}
	\end{figure}

	We now examine the mean free photon number $\langle p\rangle$, which is the sum of the populations of all states in the Hilbert space multiplied by the number of free photons they carry. This macroscopic quantity integrates the contributions from all energy levels and provides a global measure of the system's photonic state.
	
	\subsubsection{Re-entrant dynamics of the mean photon number}
	\label{subsubsec:Re-entrant}
	
	\begin{figure}
		\centering
		\includegraphics[width=.5\textwidth]{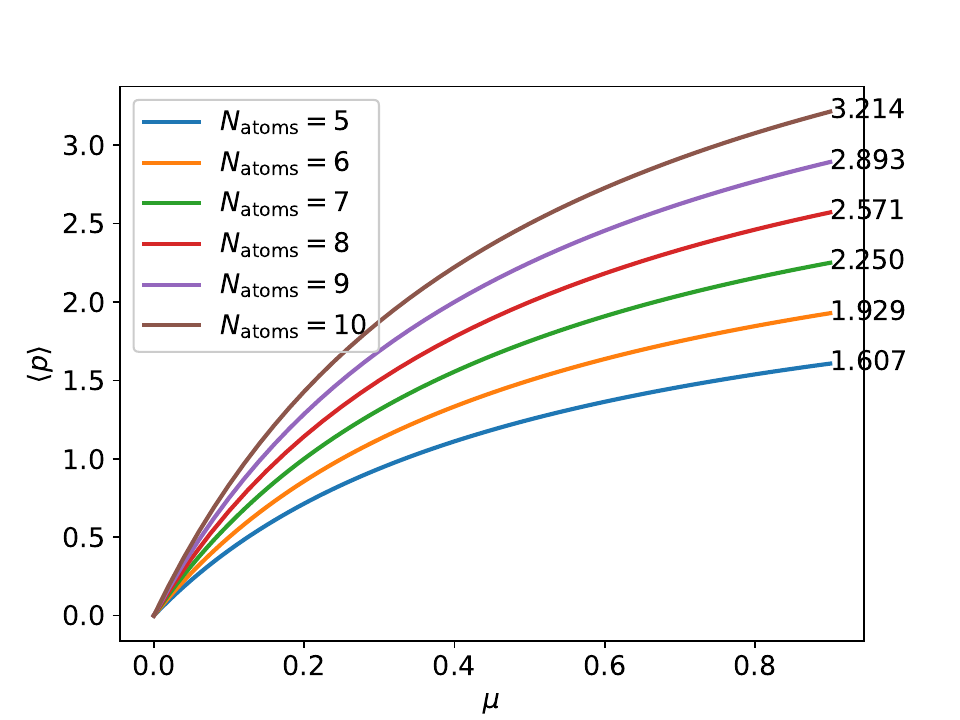} 
		\caption{(online color) {\it Steady-state mean free photon number versus pumping rate $\mu$ for $N_\mathrm{atoms}=5$ to $10$.} $\langle p\rangle$ increases monotonically with $\mu$ with decreasing slope (gain saturation). At $\mu = 0.9$, the values scale almost perfectly linearly with $N_\mathrm{atoms}$: $\langle p\rangle \approx 0.3214\times N_\mathrm{atoms}$, indicating independent atomic contributions.}
		\label{fig:AveragePhotonNumber-Mu}
	\end{figure}
	
	Fig.~\ref{fig:AveragePhotonNumber-T} shows the time evolution of $\langle p \rangle$ for system sizes $N_{\mathrm{atoms}} = 5$ to $10$ (panels (a)--(f)) and pumping rates $\mu = 0, 0.1, 0.3, 0.5, 0.7, 0.9$. The dynamics exhibit several distinct regimes:
	\begin{itemize}
		\item Small systems ($N_{\mathrm{atoms}} \leq 6$): For all $\mu$, each curve rises rapidly from zero to a peak, then decays monotonically to a steady state. This behavior is typical of a single-stage relaxation process.
		\item Large systems with strong pumping ($N_{\mathrm{atoms}} \geq 7$, $\mu = 0.9$): A striking re-entrant behavior emerges. After the initial rise and decay, $\langle p \rangle$ slowly increases again. For $N_{\mathrm{atoms}} \geq 9$, the steady-state value exceeds the initial peak --- a counterintuitive phenomenon where the system ends up with more photons than at its first maximum.
		\item Threshold behavior for $\mu = 0.7$: A similar but weaker re-entrant behavior appears for $\mu = 0.7$ when $N_{\mathrm{atoms}} \geq 8$, indicating that both system size and pumping rate must exceed certain thresholds to observe this phenomenon.
	\end{itemize}

	The re-entrant dynamics arise from the pump-driven redistribution of population to higher energy levels, as illustrated in Figs.~\ref{fig:Dissipative+Pumped-5Atoms} and \ref{fig:Dissipative+Pumped-10Atoms}. The time evolution can be understood in three stages:
	\begin{enumerate}
		\item Initial photon burst (Stage I): Initially, all atoms are excited and no photons are present. As atoms decay, photons are rapidly emitted, causing $\langle p \rangle$ to rise sharply to a peak.
		\item Dip (Stage II): The initial burst depletes the excited atoms, and decay dominates over pumping. Photons leak out of the cavity faster than they are replenished, causing $\langle p \rangle$ to drop.
		\item Re-entrant rise (Stage III): Pumping continues to inject photons into the cavity. Some of these injected photons are absorbed by ground-state atoms, exciting them. These excited atoms subsequently decay, emitting new photons. In large systems with strong pumping, the injection of photon raises $\langle p \rangle$ above the level reached in Stage II. For $N_{\mathrm{atoms}} \geq 9$ with $\mu = 0.9$, the steady-state value even surpasses the initial peak. This is possible because the initial peak corresponds to a distribution where many excitations are still in atomic form (contributing little to $\langle p \rangle$), while the steady-state distribution maintains a higher proportion of excitations in photonic form (contributing directly to $\langle p \rangle$), without exceeding the total excitation limit $N_{\mathrm{atoms}}$.
	\end{enumerate}

	This re-entrant behavior is a hallmark of the nonlinear competition between decay and pumping in a system with multiple energy levels. The existence of clear thresholds in both $N_{\mathrm{atoms}}$ (size) and $\mu$ (pumping strength) indicates that this is not a trivial linear effect but rather an emergent phenomenon requiring sufficient system complexity.
	
	\subsubsection{Steady-state scaling with pumping rate}
	\label{subsubsec:Steady-stateScaling}
	
	Fig.~\ref{fig:AveragePhotonNumber-Mu} shows the steady-state mean photon number as a function of the pumping rate $\mu$ for $N_{\mathrm{atoms}} = 5$ to $10$. Several features are noteworthy:

	\begin{itemize}
		\item Monotonic increase: $\langle p \rangle$ increases monotonically with $\mu$ for all system sizes, as expected: stronger pumping leads to more photons.
		\item Gain saturation: The slope $d\langle p \rangle/d\mu$ decreases as $\mu$ increases, indicating that the system approaches saturation where additional pumping yields diminishing returns.
		\item Linear scaling with $N_{\mathrm{atoms}}$: At $\mu = 0.9$, the steady-state photon numbers are $1.607, 1.929, 2.250, 2.571, 2.893, 3.214$ for $N_{\mathrm{atoms}} = 5$ to $10$, respectively. These values scale almost perfectly linearly: $\langle p \rangle \approx 0.3214 \times N_{\mathrm{atoms}}$.
	\end{itemize}

	The linear scaling $\langle p \rangle \propto N_{\mathrm{atoms}}$ at $\mu = 0.9$ suggests that each atom contributes independently to the steady-state photon number, consistent with the interpretation that atoms do not share photons. However, the re-entrant dynamics observed in Fig.~\ref{fig:AveragePhotonNumber-T} are not a simple linear superposition of independent atomic behaviors; they require the system to have enough energy levels (i.e., sufficiently large $N_{\mathrm{atoms}}$) to support the pump-driven redistribution process. This interplay between linear steady-state scaling and nonlinear dynamical thresholds is a central result of this work.

	The re-entrant behavior --- specifically, the steady-state photon number exceeding the initial peak --- arises from a two-stage process. First, the initial decay of excited atoms produces a rapid photon burst (the initial peak). Second, after this burst subsides, pumping continues to inject photons into the cavity. Some of these injected photons are absorbed by ground-state atoms, exciting them, and these excited atoms subsequently decay, emitting new photons. In large systems with strong pumping, this injection can generate a second wave of photons, pushing the steady-state photon number above the initial peak. This two-stage process is absent in small systems or under weak pumping, where the pump cannot sufficiently sustain this cycle after the initial burst.
	
	\begin{figure}
		\centering
		\includegraphics[width=1.\textwidth]{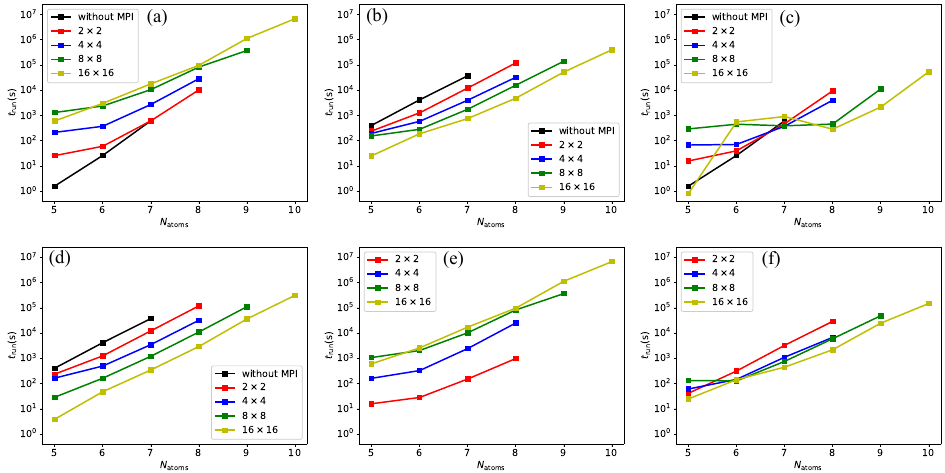} 
		\caption{(online color) {\it Comparison of time costs for solving unitary and non-unitary terms using different processor grids.} Panels (a.1)--(a.3) show the time cost of solving unitary terms: (a.1) is the total time, (a.2) the time for multiply-accumulate operations, and (a.3) the time for cross-processor communication. Panels (b.1)--(b.3) show the corresponding costs for non-unitary terms: (b.1) the total time, (b.2) the time for multiply-accumulate operations, and (b.3) the time for cross-processor communication. The horizontal axis represents the number of atoms.}
		\label{fig:ComparisonTime}
	\end{figure}
	
	\subsection{Efficiency analysis}
	\label{subsec:EffAnalysis}
	
	\begin{figure}
		\centering
        \includegraphics[width=1.\textwidth]{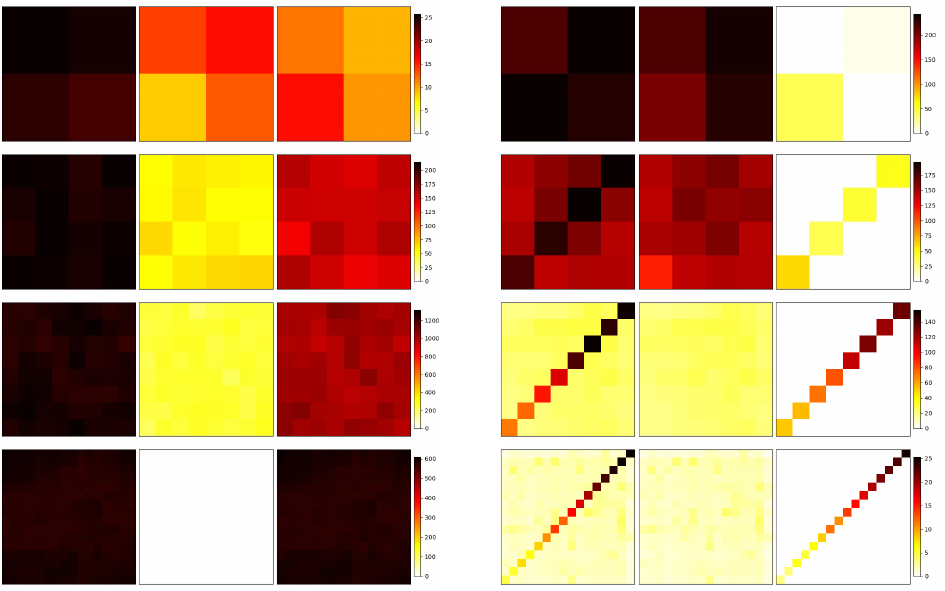}
        \caption{(online color) {\it Breakdown of computational time for the unitary (left) and non-unitary (right) terms across processor grids of varying sizes for the 5-atom case.} Each row corresponds to a different processor grid configuration: from top to bottom, $2\times 2$, $4\times 4$, $8\times 8$, and $16\times 16$ processors. For each term, three columns represent different types of measured time per processor: the left column within each panel shows the total time; the middle column shows the time spent on multiply-accumulate operations; and the right column shows the time spent on inter-processor communication. The color intensity in each heatmap represents the time required for each processor to complete the task in 10 iterations; the darker the color, the longer the time. This combined figure enables a direct comparison between the unitary and non-unitary terms, highlighting their distinct computational patterns and communication overheads as the grid size increases.}
        \label{fig:Heatmap}
	\end{figure}
	
	We solved the Lindblad master equations on a supercomputer platform. We designed four processor grids, partitioning each matrix into $2\times 2=4$, $4\times 4=16$, $8\times 8=64$, and $16\times 16=256$ blocks, respectively. We compared the time cost with and without distributed computing, and the results are shown in Fig.~\ref{fig:ComparisonTime}. First, we found that the time cost of computing unitary terms did not decrease as the number of processors increased (see panel (a.1)). Comparing panels (a.2) and (a.3), we found that while increasing the number of processors reduces the multiply-accumulate operations per processor (see panel (a.2)), it also increases cross-processor communication, leading to a rapid rise in communication costs (see panel (a.3)). Ultimately, this results in the inefficiency of distributed computing for unitary terms. Conversely, distributed computation is highly efficient for non-unitary terms (see panel (b.1)). This is because allocating more processors not only reduces multiply-accumulate operations per processor, improving efficiency (see panel (b.2)), but also because each dissipative channel involves only a single point operation that possibly requires cross-processor communication, resulting in extremely low communication costs and minimal congestion. Solving unitary terms, however, requires cross-processor block transfers, which can easily lead to congestion. Therefore, we conclude that distributed computation of the QME does not significantly improve the efficiency of solving unitary terms, but greatly improves the efficiency of solving non-unitary terms. This is particularly beneficial for large-dimensional complex quantum systems, where the number of dissipative channels $M$ is often much larger than the Hamiltonian dimension $N$, meaning that solving non-unitary terms dominates the overall computational cost. The distributed computation method proposed in this paper significantly improves the efficiency of solving non-unitary terms, making it a powerful tool for simulating open quantum systems.
		
	We now examine the breakdown of time cost per processor for the case $N_\mathrm{atoms}=5$. As shown in the left part of Fig.~\ref{fig:Heatmap}, when solving unitary terms, the total time cost increases significantly with the number of processors. The proportion of communication time relative to the total time per processor increases steadily, becoming almost equal to the total time when the number of processors reaches 256. Furthermore, the multiply-accumulate computation and communication times for unitary terms are similar across all processors, indicating that the workload is well-balanced and all processors are utilized efficiently. As shown in the right part of Fig.~\ref{fig:Heatmap}, when solving non-unitary terms, the total time cost decreases significantly as the number of processors increases. Although the proportion of communication time also grows, it remains much lower than in the unitary case, and the overall efficiency improves substantially. In addition, we observe that for non-unitary terms, cross-processor communication occurs only on diagonal processors, while off-diagonal processors remain idle during these operations. However, as noted earlier, cross-processor communication for each dissipation channel involves only a single-point operation, so this imbalance has little impact on the overall efficiency gain. For a detailed breakdown of the computation time for unitary and non-unitary terms across different processor grid sizes for systems with 5 to 10 atoms, please refer to the Appendix~\ref{appxsec:Breakdown}.
	
	\section{Conclusion and outlook}
	\label{sec:Conclusion}

	In this work, we have developed a distributed computing framework for solving the Lindblad master equation in large-dimensional cavity QED systems, specifically the dissipative-pumped Tavis--Cummings model. By leveraging the sparsity of jump operators $\hat{A}_k = |j\rangle\langle i|$ and combining it with Cannon's algorithm, we significantly improve the computational efficiency of non-unitary evolution.

	For unitary terms, we employ a Taylor series approximation combined with Cannon's algorithm to handle matrix exponentials. Our results show that while increasing the number of processors reduces the multiply-accumulate computation cost per processor, the cross-processor communication overhead increases substantially, limiting scalability. When using 256 processors, the communication time becomes comparable to the total computation time. Nevertheless, this method provides a feasible distributed solution for large-dimensional quantum systems that cannot be simulated on a single processor.

	For non-unitary terms, we exploit the sparsity of the jump operators to simplify matrix operations into three basic updates: a point operation (population transfer), a row operation, and a column operation. This reduces the computational complexity from $\mathcal{O}(MN^3)$ to $\mathcal{O}(MN)$. Distributed computation for non-unitary terms demonstrates high efficiency: the total computation time decreases significantly with increasing processor count, and communication overhead remains extremely low, as only diagonal processors may require single-point communication.
	
	The dynamic subspace construction method further reduces the Hamiltonian dimension to $5.63\%$ of the full dimension when $N_{\mathrm{atoms}} = 10$, with memory usage only $0.32\%$ of the full Hamiltonian, enabling simulations of large-scale open quantum systems.

	Beyond the numerical method, we have uncovered several nonlinear phenomena in the dissipative-pumped Tavis--Cummings model under global excitation constraints (hard-core boson constraint, total excitations limited to the initial value $N_{\mathrm{atoms}}E$):
	\begin{itemize}
		\item Non-monotonic plateau: The steady-state peak population $\rho_{\mathrm{peak}}$ does not decay monotonically with energy level. Instead, a plateau (or weak increase) emerges at intermediate energy levels, where multiple levels achieve comparable maximum populations under their respective optimal pumping rates. As $N_{\mathrm{atoms}}$ increases, this plateau shifts to higher energy levels and broadens, with the plateau width growing monotonically from $W=2$ ($N_{\mathrm{atoms}}=5$) to $W=4$ ($N_{\mathrm{atoms}}=10$).
		\item Re-entrant dynamics: The mean free photon number $\langle p \rangle$ exhibits re-entrant behavior in large systems ($N_{\mathrm{atoms}} \geq 7$) under strong pumping ($\mu = 0.9$): after an initial decay, $\langle p \rangle$ slowly increases again, and for $N_{\mathrm{atoms}} \geq 9$, the steady-state value exceeds the initial peak. This counterintuitive phenomenon arises from pump-driven photon injection, subsequent absorption by ground-state atoms, and re-emission of new photons — a cycle that requires sufficient system size and pumping strength.
		\item Threshold behavior: Clear thresholds exist in both system size and pumping rate. Re-entrant dynamics appear for $\mu = 0.9$ when $N_{\mathrm{atoms}} \geq 7$, and for $\mu = 0.7$ when $N_{\mathrm{atoms}} \geq 8$, indicating that this is an emergent phenomenon requiring sufficient system complexity.
		\item Linear steady-state scaling: At $\mu = 0.9$, the steady-state photon number scales linearly with $N_{\mathrm{atoms}}$: $\langle p \rangle \approx 0.3214 \times N_{\mathrm{atoms}}$, indicating that each atom contributes independently in the steady state, despite the nonlinear dynamical thresholds.
	\end{itemize}

	Future work will focus on extending the proposed method to other open quantum system models, optimizing load balancing for non-unitary computations, and exploring applications in quantum information processing.

	\begin{acknowledgments}
		The reported study was funded by China Scholarship Council, project number 202108090483. The authors acknowledge Center for Collective Usage of Ultra HPC Resources (https://parallel.ru/) at Lomonosov Moscow State University for providing supercomputer resources that have contributed to the research results reported within this paper.
	\end{acknowledgments}
	
\appendix

	\section{Method for constructing the dynamic subspace}
	\label{appxsec:DynamicSubspace}
	
	\subsection{Dynamic path diagram: an example of the TCM with two two-level atoms}
	\label{appxsubsec:DynamicPath}
	
	The traditional tensor product is commonly used to construct Hamiltonians, as it directly builds upon expressions composed of various operators. However, this method has a significant drawback: the resulting Hamiltonian becomes extremely large, especially in systems with many degrees of freedom, as it contains numerous redundant states that do not participate in the evolution. To address this issue, we introduce a dynamic subspace construction method that derives the entire dynamic path diagram from an initial state and significantly reduces computational costs. The method comprises the following three steps:
	\begin{itemize}
		\item Based on the initial state and all possible interactions and dissipation channels relevant to the quantum dynamics, generate and label all states that may participate in the evolution process.
		\item Construct a dynamical path graph that includes all generated states.
		\item Use this graph to construct a Hamiltonian and a set of dissipation channels that cover the aforementioned states along with their interactions and dissipation mechanisms.
	\end{itemize}
	
\begin{figure}
		\centering
        \includegraphics[width=.7\textwidth]{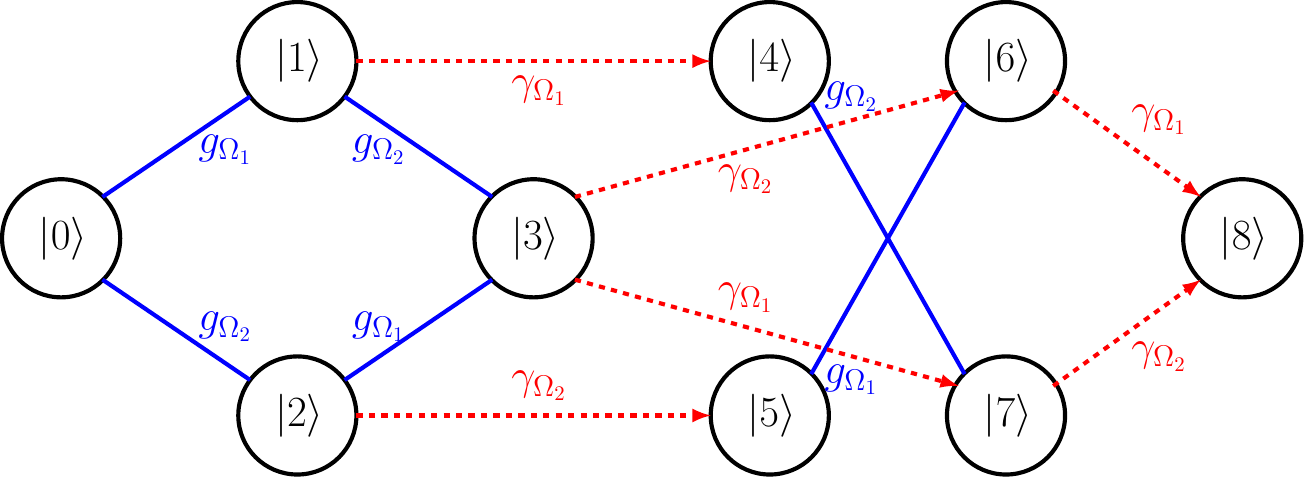}
        \caption{(online color) {\it Dynamic path diagram for the TCM with two two-level atoms.} Here, $g_{\Omega_1}$ and $\gamma_{\Omega_1}$ denote the interaction strength and dissipative rate of the first atom, respectively, while $g_{\Omega_2}$ and $\gamma_{\Omega_2}$ denote those of the second atom. The solid blue line between two quantum states represents an interaction, while the red arrow represents a dissipation channel. The direction of the arrow indicates the direction of dissipation.}
        \label{appxfig:DynamicPath}
	\end{figure}

	Next, we take a TCM with two two-level atoms as an example.  A single two-level atom coupled to a photonic mode has a Hilbert space of dimension $2^2=4$, consisting of the four quantum states: $|0\rangle_{\text{ph}_1}|0\rangle_{\text{at}_1}$, $|0\rangle_{\text{ph}_1}|1\rangle_{\text{at}_1}$, $|1\rangle_{\text{ph}_1}|0\rangle_{\text{at}_1}$, and $|1\rangle_{\text{ph}_1}|1\rangle_{\text{at}_1}$. Suppose the initial state is an excited atom with no photons, i.e., $|\Psi_{\text{initial}}\rangle=|0\rangle_{\text{ph}_1}|1\rangle_{\text{at}_1}$. Starting from this initial state, only two states are reachable via interactions and dissipation channels: $|0\rangle_{\text{ph}_1}|0\rangle_{\text{at}_1}$ and $|1\rangle_{\text{ph}_1}|0\rangle_{\text{at}_1}$. The state $|1\rangle_{\text{ph}_1}|1\rangle_{\text{at}_1}$ is not accessible. Thus, this state can be eliminated, yielding a reduced Hilbert space of dimension 3. Similarly, if the system contains two excited atoms and no photons, the Hilbert space dimension can be reduced from $2^4=16$ to $3^2=9$. In this case, the nine accessible quantum states are as follows:
	\begin{equation}
		\begin{aligned}
			|0\rangle&=|0\rangle_{\text{ph}_1}|1\rangle_{\text{at}_1}|0\rangle_{\text{ph}_2}|1\rangle_{\text{at}_2},\\
			|1\rangle&=|1\rangle_{\text{ph}_1}|0\rangle_{\text{at}_1}|0\rangle_{\text{ph}_2}|1\rangle_{\text{at}_2},\\
			|2\rangle&=|0\rangle_{\text{ph}_1}|1\rangle_{\text{at}_1}|1\rangle_{\text{ph}_2}|0\rangle_{\text{at}_2},\\
			|3\rangle&=|1\rangle_{\text{ph}_1}|0\rangle_{\text{at}_1}|1\rangle_{\text{ph}_2}|0\rangle_{\text{at}_2},\\
			|4\rangle&=|0\rangle_{\text{ph}_1}|0\rangle_{\text{at}_1}|0\rangle_{\text{ph}_2}|1\rangle_{\text{at}_2},\\
			|5\rangle&=|0\rangle_{\text{ph}_1}|1\rangle_{\text{at}_1}|0\rangle_{\text{ph}_2}|0\rangle_{\text{at}_2},\\
			|6\rangle&=|1\rangle_{\text{ph}_1}|0\rangle_{\text{at}_1}|0\rangle_{\text{ph}_2}|0\rangle_{\text{at}_2},\\
			|7\rangle&=|0\rangle_{\text{ph}_1}|0\rangle_{\text{at}_1}|1\rangle_{\text{ph}_2}|0\rangle_{\text{at}_2},\\
			|8\rangle&=|0\rangle_{\text{ph}_1}|0\rangle_{\text{at}_1}|0\rangle_{\text{ph}_2}|0\rangle_{\text{at}_2}. 
		\end{aligned}
	\end{equation}		
	The dynamic path diagram for the TCM with two two-level atoms is shown in Fig.~\ref{appxfig:DynamicPath}.
	
	\subsection{Efficiency analysis}
	\label{appxsubsec:Efficiency}
	
	The core advantage of this method lies in its ability to directly eliminate redundant quantum states during the Hamiltonian construction stage. Let $\mathcal{C}_1$ denote the complete Hilbert space generated by the traditional tensor product method, and $\hat{H}_1$ the corresponding Hamiltonian. Let $\mathcal{C}_2$ denote the reduced space obtained from the dynamic subspace construction method proposed in this paper, and $\hat{H}_2$ the corresponding Hamiltonian. Then we have $\mathcal{C}_2\subset\mathcal{C}_1$ and $\dim\left(\hat{H}_2\right)\leq\dim\left(\hat{H}_1\right)$. As the number of atoms increases, the dimension of the complete Hamiltonian $\hat{H}_1$ grows exponentially with base 4. In contrast, the dimension of the reduced Hamiltonian $\hat{H}_2$ also grows exponentially, but with base 3. Therefore, as the number of atoms increases, the dimension of the latter becomes much smaller than that of the former. Table~\ref{appxtab:Efficiency} compares the dimensions of Hamiltonians constructed by the two methods for atom numbers ranging from 1 to 10, and lists their ratios. It also provides the ratio of memory usage for Hamiltonians constructed by the two methods. In addition, the interaction dimension $\dim(\mathcal{I})$ and the dissipation channel dimension $\dim(\mathcal{K})$ are listed. As shown in the fifth row of the table, the ratio of the memory overhead of the dynamic subspace method to that of the traditional method decreases as the number of atoms increases. When $N_\mathrm{atoms}=4$, this ratio reaches 0.1, and when $N_\mathrm{atoms}=8$, it reaches 0.01. This greatly reduces memory overhead and improves computational efficiency. Furthermore, we see from the table that for $N_\mathrm{atoms}\ge 4$, $\dim\left(\mathcal{K}\right)$ exceeds $\dim\left(\hat{H}_2\right)$, and the difference grows rapidly with increasing atom number. This indicates that after applying the dynamic subspace reduction, the number of dissipation channels $M$ (denoted by $\dim(\mathcal{K})$ in the table) becomes much larger than the reduced Hamiltonian dimension $N=\dim\left(\hat{H}_2\right)$ for large atom numbers.
	\begin{table}[!htpb]
        \centering
		\begin{tabular}{|c|c|c|c|c|c|c|c|c|c|c|c|c|c|}
			\hline
			Number of atoms & 1 & 2 & 3 & 4 & 5 & 6 & 7 & 8 & 9 & 10 & $\dots$ & $N_\mathrm{atoms}$ \\
			\hline
			$\dim\left(\hat{H}_1\right)$ & 4 & 16 & 64 & 256 & 1024 & 4096 & 16384 & 65536 & 262144 & 1048576 & $\dots$ & $4^{N_\mathrm{atoms}}$ \\
			\hline
			$\dim\left(\hat{H}_2\right)$ & 3 & 9 & 27 & 81 & 243 & 729 & 2187 & 6561 & 19683 & 59049 & $\dots$ & $3^{N_\mathrm{atoms}}$ \\
			\hline
			$\dfrac{\dim\left(\hat{H}_2\right)}{\dim\left(\hat{H}_1\right)}$ & 0.75 & 0.5625 & 0.4219 & 0.3164 & 0.2373 & 0.178 & 0.1335 & 0.1001 & 0.0751 & 0.0563 & $\dots$ & $\left(\dfrac{3}{4}\right)^{N_\mathrm{atoms}}$ \\
			\hline
			$\dfrac{\dim^2\left(\hat{H}_2\right)}{\dim^2\left(\hat{H}_1\right)}$ & 0.5625 & 0.3164 & 0.178 & 0.1001 & 0.0563 & 0.0317 & 0.0178 & 0.0100 & 0.0056 & 0.0032 & $\dots$ & $\left(\dfrac{3}{4}\right)^{2N_\mathrm{atoms}}$ \\
			\hline
			$\dim\left(\mathcal{I}\right)$ & 1 & 6 & 27 & 108 & 405 & 1458 & 5103 & 17496 & 59049 & 196830 & $\dots$ & $N_\mathrm{atoms}\times3^{N_\mathrm{atoms}-1}$ \\
			\hline
			$\dim\left(\mathcal{K}\right)$ & 1 & 6 & 27 & 108 & 405 & 1458 & 5103 & 17496 & 59049 & 196830 & $\dots$ & $N_\mathrm{atoms}\times3^{N_\mathrm{atoms}-1}$ \\
			\hline
		\end{tabular}
		\caption{{\it Comparison of Hamiltonian dimensions constructed by the two methods for different numbers of atoms.} This table compares the dimensions of the full Hamiltonian $\hat{H}_1$, constructed using the traditional tensor product method, and the reduced Hamiltonian $\hat{H}_2$, constructed using the dynamic subspace method, for atom numbers ranging from $n = 1$ to $10$. The fourth and fifth rows show the ratio of the Hamiltonian dimensions obtained by the two methods and the ratio of the memory required to construct the Hamiltonians, respectively. It also lists the corresponding interaction set dimension $\dim\left(\mathcal{I}\right)$ and dissipation channel dimension $\dim\left(\mathcal{K}\right)$.}	
		\label{appxtab:Efficiency}
	\end{table}
	
	\section{Distributed implementation of the non-unitary dissipation}
	\label{appxsec:Non-unitary}
	
	We consider a quantum system of dimension $N$. The density matrix $\hat{\rho}$ is an $N\times N$ Hermitian matrix, with matrix elements $\hat{\rho}_{m,n}=\langle m|\hat{\rho}|n\rangle$. For a specific dissipation channel corresponding to the transition from state $|i\rangle$ to state $|j\rangle$, the associated jump operator is $\hat{A}_k=|j\rangle\langle i|$. In matrix form, this operator has a $1$ at the $j$-th row and $i$-th column, and $0$ elsewhere. The complexity of the dissipative term in the Lindblad master equation is as follows:
	\begin{equation}
		\label{appxeq:DissSuperComplexity}
		\hat{L}_k\left(\hat{\rho}\right)=\underbrace{\gamma_k\left(\underbrace{\hat{A}_k\hat{\rho}\hat{A}_k^\dagger}_{\mathcal{O}(N^3)}-\frac{1}{2}\left(\underbrace{\hat{\rho}\hat{A}_k^\dagger\hat{A}_k}_{\mathcal{O}(N^3)}+\underbrace{\hat{A}_k^\dagger\hat{A}_k\hat{\rho}}_{\mathcal{O}(N^3)}\right)\right)}_{\mathcal{O}(N^3)},
	\end{equation}
	Each channel requires $\mathcal{O}(N^3)$ operations. Assuming there are $M$ dissipation channels, the total complexity is:
	\begin{equation}
		\label{appxeq:LindbladOperatorComplexity}
		\hat{L}\left(\hat{\rho}\right)=\underbrace{\sum_{k\in \mathcal{K}}\underbrace{\hat{L}_k\left(\hat{\rho}\right)}_{\mathcal{O}(N^3)}}_{\mathcal{O}(MN^3)}.
	\end{equation}
	The total complexity is $\mathcal{O}(MN^3)$, which does not scale well with the number of atoms. However, each $\hat{A}_k$ is extremely sparse, containing only one non-zero element, and we can exploit this sparsity.
	
	\subsection{Simplified update for the term $\hat{A}_k\hat{\rho}\hat{A}_k^\dagger$}
	\label{appxsubsec:SimplifiedTerm1}

	We first compute the matrix elements of $\hat{A}_k\hat{\rho}$. The notation $(\hat{A}_k\hat{\rho})_{m,n}$ denotes the element at the $m$-th row and $n$-th column of the matrix product $\hat{A}_k\hat{\rho}$. Using the definition of matrix multiplication in the computational basis, we have:
	\begin{equation}
		\label{appxeq:FirstTermLeftElement}
		(\hat{A}_k\hat{\rho})_{m,n}=\sum_p(\hat{A}_k)_{m,p}\hat{\rho}_{p,n}
	\end{equation}

	For convenience, we introduce the Kronecker delta function, defined as:
	\begin{equation}
		\label{appxeq:Kronecker}
		\delta_{a,b}=\begin{cases} 1, & \text{if } a=b,\\ 0, & \text{otherwise}. \end{cases}
	\end{equation}
	
	Given the jump operator $\hat{A}_k=|j\rangle\langle i|$, its matrix elements are given by:
	\begin{equation}
		\label{appxeq:JumpOperatorElementKronecker}
		\begin{aligned}
			(\hat{A}_k)_{m,p}&=\langle m|j\rangle\langle i|p\rangle \\
			&=\delta_{m,j}\delta_{i,p}.
		\end{aligned}
	\end{equation}
	Substituting this into Eq.~\eqref{appxeq:FirstTermLeftElement} yields:
	\begin{equation}
		\label{appxeq:FirstTermLeftElementKronecker}
		\begin{aligned}
			(\hat{A}_k\hat{\rho})_{m,n}&=\sum_p\delta_{m,j}\delta_{i,p}\hat{\rho}_{p,n}\\
			&=\delta_{m,j}\hat{\rho}_{i,n}.
		\end{aligned}
	\end{equation}
	From this result, we observe that:
	\begin{itemize}
		\item Only the $j$-th row of the resulting matrix is non-zero.
    		\item This non-zero row is identical to the $i$-th row of the original density matrix $\hat{\rho}$.
	\end{itemize}
	
	We now multiply the result from the previous step on the right by $\hat{A}_k^\dagger$:
	\begin{equation}
		\label{appxeq:FirstTermRightElement}
		(\hat{A}_k\hat{\rho}\hat{A}_k^\dagger)_{p,q}=\sum_m(\hat{A}_k\hat{\rho})_{p,m}(\hat{A}_k^\dagger)_{m,q}.
	\end{equation}
	The matrix elements of $\hat{A}_k^\dagger=|i\rangle\langle j|$ are given by:
	\begin{equation}
		\label{appxeq:JumpOperatorConjugateElementKronecker}
		\begin{aligned}
			(\hat{A}_k^\dagger)_{m,q}&=\langle m|i\rangle\langle j|q\rangle\\
			&=\delta_{m,i}\delta_{j,q}.
		\end{aligned}
	\end{equation}

	Substituting the expression $(\hat{A}_k\rho)_{p,m}=\delta_{p,j}\hat{\rho}_{i,m}$ obtained above, we obtain:
	\begin{equation}
		\label{appxeq:FirstTermRightElementKronecker}
		\begin{aligned}
			(\hat{A}_k\hat{\rho}\hat{A}_k^\dagger)_{p,q}&=\sum_m\delta_{p,j}\hat{\rho}_{i,m}\delta_{m,i}\delta_{j,q}\\
			&=\delta_{p,j}\hat{\rho}_{i,i}\delta_{j,q}
		\end{aligned}
	\end{equation}
	
	Combining the results from the previous steps, we obtain:
	\begin{equation}
		\label{appxeq:FirstTermRight}
		\hat{A}_k\hat{\rho}\hat{A}_k^\dagger=\hat{\rho}_{i,i}|j\rangle\langle j|.
	\end{equation}

	That is, in matrix form:
	\begin{itemize}
   		\item Only the $(j,j)$ element is non-zero.
    		\item Its value is $\hat{\rho}_{i,i}$.
	\end{itemize}

	Operation: Take $\hat{\rho}_{i,i}$ and add it to $\hat{\rho}_{j,j}$.

	Computational complexity: $\mathcal{O}(1)$.
	
	\subsection{Simplified update for the term $-\dfrac{1}{2}\{\hat{\rho},\hat{A}_k^\dagger\hat{A}_k\}$}
	\label{appxsubsec:SimplifiedTerm2}

	We first compute the product $\hat{A}_k^\dagger\hat{A}_k$:
	\begin{equation}
		\label{appxeq:Conjugate*JumpOperator}
		\begin{aligned}
			\hat{A}_k^\dagger\hat{A}_k&=|i\rangle\langle j|j\rangle\langle i|\\
			&=|i\rangle\langle i|.
		\end{aligned}
	\end{equation}
	That is, $\hat{A}_k^\dagger\hat{A}_k$ is the projection operator onto state $|i\rangle$. The anti-commutator expands to:
	\begin{equation}
		\label{appxeq:SecondTerm}
		\{\hat{\rho},\hat{A}_k^\dagger\hat{A}_k\}=\hat{\rho}|i\rangle\langle i|+|i\rangle\langle i|\hat{\rho}.
	\end{equation}

	The matrix elements of the first component are given by:
	\begin{equation}
		\label{appxeq:SecondTermLeftElementKronecker}
		\begin{aligned}
			(\hat{\rho}|i\rangle\langle i|)_{m,n}&=\langle m|\hat{\rho}|i\rangle\langle i|n\rangle\\
			&=\hat{\rho}_{m,i}\delta_{i,n}.
		\end{aligned}
	\end{equation}
	Therefore:
	\begin{itemize}
    		\item Only the $i$-th column of the resulting matrix is non-zero.
    		\item This non-zero column is identical to the $i$-th column of the original density matrix $\hat{\rho}$.
	\end{itemize}
	
	Similarly, the matrix elements of the second component are given by:
	\begin{equation}
		\label{appxeq:SecondTermRightElementKronecker}
		\begin{aligned}
			(|i\rangle\langle i|\hat{\rho})_{m,n}&=\langle m|i\rangle\langle i|\hat{\rho}|n\rangle\\
			&=\delta_{m,i}\hat{\rho}_{i,n}.
		\end{aligned}
	\end{equation}
	Therefore:
	\begin{itemize}
    		\item Only the $i$-th row of the resulting matrix is non-zero.
    		\item This non-zero row is identical to the $i$-th row of the original density matrix $\hat{\rho}$.
	\end{itemize}
	
	Combining the results from Eqs.~\eqref{appxeq:SecondTermLeftElementKronecker} and \eqref{appxeq:SecondTermRightElementKronecker}, the matrix elements of the full anti-commutator are given by:
	\begin{equation}
		\label{appxeq:SecondTermElement}
		\{\hat{\rho},\hat{A}_k^\dagger\hat{A}_k\}_{m,n}=
		\begin{cases}
			\hat{\rho}_{i,n}, & \text{if }m=i\text{ and }n\neq i,\\[4pt]
			2\hat{\rho}_{i,i}, & \text{if }m=n=i,\\[4pt]
			\hat{\rho}_{m,i}, & \text{if }m\neq i\text{ and }n=i,\\[4pt]
			0, & \text{otherwise}.
		\end{cases}
	\end{equation}
	Multiplying the anti-commutator by the factor $-\dfrac{1}{2}$ yields:
	\begin{equation}
		\label{appxeq:SecondTermElementConstant}
		-\frac{1}{2}\{\hat{\rho},\hat{A}_k^\dagger\hat{A}_k\}_{m,n}=
		\begin{cases}
			-\frac{1}{2}\hat{\rho}_{i,n}, & \text{if }m=i\text{ and }n\neq i,\\[4pt]
			-\hat{\rho}_{i,i}, & \text{if }m=n=i,\\[4pt]
			-\frac{1}{2}\hat{\rho}_{m,i}, & \text{if }m\neq i\text{ and }n=i,\\[4pt]
			0, & \text{otherwise}.
		\end{cases}
	\end{equation}

	The contribution of this term is implemented as the following operations on the density matrix:
	\begin{itemize}
    		\item Row $i$ (excluding the diagonal element $(i,i)$): Each off-diagonal element $\hat{\rho}_{i,n}$ (with $n\neq i$) is halved.
	    \item Column $i$ (excluding the diagonal element $(i,i)$): Each off-diagonal element $\hat{\rho}_{m,i}$ (with $m \neq i$) is halved.
	    \item Diagonal element $(i,i)$: The element $\hat{\rho}_{i,i}$ is set to zero, as it receives a contribution of $-\dfrac{1}{2}\hat{\rho}_{i,i}$ from both the row and column updates.
	\end{itemize}

	Computational complexity:
	\begin{itemize}
  		\item Operations on the $i$-th row require $\mathcal{O}(N)$ time.
   		\item Operations on the $i$-th column require $\mathcal{O}(N)$ time.
	    \item Thus, the total complexity per channel is $\mathcal{O}(N)$.
	\end{itemize}
	
	\subsection{Combined local update rules}
	\label{appxsubsec:Combined}

	We consider the parameters $\hbar$, $\gamma_k$ and $\Delta t$ in Eqs.~\eqref{eq:Non-unitaryTerm} and \eqref{appxeq:DissSuperComplexity}, and set $\hbar=1$.
	
	Effect on $\hat{\rho}_{j,j}$:
	\begin{itemize}
    		\item From the first term: $+\gamma_k\Delta t\hat{\rho}_{i,i}$.
	    \item From the second term: None, since the second term does not affect $\hat{\rho}_{j,j}$ when $j\neq i$.
	\end{itemize}
	Therefore:
	\begin{equation}
		\label{appxeq:EffectonRhojj}
		\hat{\rho}_{j,j}=\hat{\rho}_{j,j}+\gamma_k\Delta t\hat{\rho}_{i,i}.
	\end{equation}

	Effect on $\hat{\rho}_{i,i}$:
	\begin{itemize}
    		\item From the first term: None, the first term only affects $\hat{\rho}_{j,j}$.
	    \item From the second term: $-\gamma_k\Delta t\hat{\rho}_{i,i}$.
	\end{itemize}
	Therefore:
	\begin{equation}
		\label{appxeq:EffectonRhoii}
		\hat{\rho}_{i,i}=\hat{\rho}_{i,i}-\gamma_k\Delta t\hat{\rho}_{i,i}.
	\end{equation}

	Effect on other elements in row $i$ ($\hat{\rho}_{i,n}$ with $n\neq i$):
	\begin{itemize}
    		\item From the first term: None.
	    \item From the second term: $-\dfrac{\gamma_k\Delta t}{2}\hat{\rho}_{i,n}$.
	\end{itemize}
	Therefore:
	\begin{equation}
		\label{appxeq:EffectonRowi}
		\hat{\rho}_{i,n}=\hat{\rho}_{i,n}-\frac{\gamma_k\Delta t}{2}\hat{\rho}_{i,n}, \quad\forall n\neq i.
	\end{equation}

	Effect on other elements in column $i$ ($\hat{\rho}_{m,i}$ with $m\neq i$):
	\begin{itemize}
    		\item From the first term: None.
	    \item From the second term: $-\dfrac{\gamma_k\Delta t}{2}\hat{\rho}_{m,i}$.
	\end{itemize}
	Therefore:
	\begin{equation}
		\label{appxeq:EffectonColumni}
		\hat{\rho}_{m,i}=\hat{\rho}_{m,i}-\frac{\gamma_k\Delta t}{2}\hat{\rho}_{m,i}, \quad\forall m\neq i.
	\end{equation}
	
	Other elements: All remaining matrix elements are unaffected by this dissipation channel.
	
	Finally, we use Fig.~\ref{fig:Non-unitary}(a.1)--(a.3) to illustrate the effect of Eqs.~\eqref{appxeq:EffectonRhojj}--\eqref{appxeq:EffectonColumni} on the density matrix.
	
	\subsection{Pseudocode and implementation details}
	\label{appxsubsec:Pseudocode}

	For each dissipation channel $\hat{A}_k$, the numerical update is implemented as follows:
	
	\begin{algorithm}[H]
		\caption{Population transfer and decay updates for transition $i\rightarrow j$ with rate $\gamma_k$}
		\label{appxalg:DissChannel}
		\KwIn{Current density matrix $\hat{\rho}$, transition rate $\gamma_k$, time step $\Delta t$, indices $i$, $j$, system size $N$}
		\KwOut{Updated density matrix $\hat{\rho}$}

		\tcc{1. Population transfer from state $i$ to state $j$}
		$\hat{\rho}[j][j]\leftarrow\hat{\rho}[j][j]+\gamma_k\cdot\Delta t\cdot\hat{\rho}[i][i]$\;

		\tcc{2. Decay of the $i$-th row (including diagonal)}
		\For{$n\leftarrow 0$ \KwTo $N-1$}{
		    $\hat{\rho}[i][n]\leftarrow\hat{\rho}[i][n]-\left(\dfrac{\gamma_k\cdot\Delta t}{2}\right)\cdot\hat{\rho}[i][n]$\;
		}

		\tcc{3. Decay of the $i$-th column (including diagonal)}
		\For{$m\leftarrow 0$ \KwTo $N-1$}{
		    $\hat{\rho}[m][i]\leftarrow\hat{\rho}[m][i]-\left(\dfrac{\gamma_k\cdot\Delta t}{2}\right)\cdot\hat{\rho}[m][i]$\;
		}
	\end{algorithm}

	Remark: The diagonal element $\hat{\rho}_{i,i}$ is reduced by a total of $\gamma_k\Delta t\hat{\rho}_{i,i}$---once from the row decay and once from the column decay---so no additional operation is required beyond Steps 2 and 3.
	
	In Algorithm~\ref{appxalg:DissChannel}, if Step 2 is executed first, the value of $\hat{\rho}_{i,i}$ decreases during the row decay in Step 2 as:
	\begin{equation}
		\label{appxeq:Step2}
		\hat{\rho}_{i,i}\leftarrow\hat{\rho}_{i,i}-\frac{\gamma_k\Delta t}{2}\hat{\rho}_{i,i}=\hat{\rho}_{i,i}\left(1-\frac{\gamma_k\Delta t}{2}\right).
	\end{equation}
	Consequently, when Step 3 is executed, the value of $\hat{\rho}_{i,i}$ has already become $\hat{\rho}_{i,i}\left(1-\frac{\gamma_k\Delta t}{2}\right)$. The column decay in Step 3 then reduces this value by:
	\begin{equation}
		\label{appxeq:Step3}
		\frac{\gamma_k\Delta t}{2}\hat{\rho}_{i,i}\left(1-\frac{\gamma_k\Delta t}{2}\right).
	\end{equation}
	As a result, the total reduction of $\hat{\rho}_{i,i}$ is:
	\begin{equation}
		\label{appxeq:Error}
		\frac{\gamma_k\Delta t}{2}\hat{\rho}_{i,i}+\frac{\gamma_k\Delta t}{2}\hat{\rho}_{i,i}\left(1-\frac{\gamma_k\Delta t}{2}\right)=\gamma_k\Delta t\hat{\rho}_{i,i} -\frac{(\gamma_k\Delta t)^2}{4}\hat{\rho}_{i,i},
	\end{equation}
	which introduces an extra error term of order $\mathcal{O}((\Delta t)^2)$. In the limit $\Delta t \to 0$, this error term can be neglected. If we perform Step 3 first, followed by Step 2, the resulting error term will be the same.
	
	To prevent the gradual accumulation of the error term over multiple iterations, a more rigorous approach is to use a copy of the density matrix from the previous time step. Let $\hat{\rho}_\mathrm{old}=\hat{\rho}$ denote the input density matrix. All updates are then performed based on $\hat{\rho}_\mathrm{old}$ and written to $\hat{\rho}$:
	
	\begin{algorithm}[H]
		\caption{Population transfer and decay updates using old density matrix values}
		\label{appxalg:DissChannelCopy}
		\KwIn{Current density matrix $\hat{\rho}$, transition rate $\gamma_k$, time step $\Delta t$, indices $i$, $j$, system size $N$}
		\KwOut{Updated density matrix $\hat{\rho}$}

		\tcc{Make a copy of the current density matrix}
		$\hat{\rho}_\mathrm{old}\leftarrow\hat{\rho}$\;

		\tcc{1. Population transfer (using old values)}
		$\hat{\rho}[j][j]\leftarrow\hat{\rho}[j][j]+\gamma_k\cdot\Delta t\cdot\hat{\rho}_\mathrm{old}[i][i]$\;

		\tcc{2. Decay of the $i$-th row (using old values)}
		\For{$n\leftarrow 0$ \KwTo $N-1$}{
		    $\hat{\rho}[i][n]\leftarrow\hat{\rho}[i][n]-\left(\frac{\gamma_k\cdot\Delta t}{2}\right)\cdot\hat{\rho}_\mathrm{old}[i][n]$\;
		}

		\tcc{3. Decay of the $i$-th column (using old values)}
		\For{$m\leftarrow 0$ \KwTo $N-1$}{
		    $\hat{\rho}[m][i]\leftarrow\hat{\rho}[m][i]-\left(\frac{\gamma_k\cdot\Delta t}{2}\right)\cdot\hat{\rho}_\mathrm{old}[m][i]$\;
		}
	\end{algorithm}
    
    This approach ensures that all operations strictly follow the Euler integration scheme in Eq.~\eqref{eq:Non-unitaryTerm}, independent of the order in which they are applied. The additional memory cost of storing a copy of the $N\times N$ density matrix is $\mathcal{O}(N^2)$, which is acceptable for most practical simulations.
    
    \subsection{Complexity analysis}
    \label{appxsubsec:Complexity}

	The computational complexity of the two approaches is summarized in Table~\ref{appxtab:ComplexityBreakdown}:
	\begin{table}[h]
		\centering
		\begin{tabular}{|l|c|}
			\hline
			Operation & Complexity \\
			\hline
			Dissipation channel population transfer & $\mathcal{O}(1)$ \\
			Dissipation channel row decay & $\mathcal{O}(N)$ \\
			Dissipation channel column decay & $\mathcal{O}(N)$ \\
			\hline
			Total per channel & $\mathcal{O}(2N+1)=\mathcal{O}(N)$ \\
			\hline
		\end{tabular}
		\caption{{\it Computational complexity breakdown for a dissipation channel using the optimized row/column operation approach.}}
		\label{appxtab:ComplexityBreakdown}
	\end{table}
	
	We assume the number of dissipation channels is $M$:
	\begin{table}[h]
		\centering
		\begin{tabular}{|c|c|c|}
			\hline
			Method & Per Channel & $M$ Channels \\
			\hline
			Direct Matrix Multiplication & $\mathcal{O}(N^3)$ & $\mathcal{O}(MN^3)$ \\
			\hline
			Optimized (Row + Column Operations) & $\mathcal{O}(N)$ & $\mathcal{O}(MN)$ \\
			\hline
		\end{tabular}
		\caption{{\it Comparison of computational complexity between direct matrix multiplication and the optimized row/column operation approach.}}
		\label{appxtab:Complexity}
	\end{table}

	When the dimension of the quantum system and the number of dissipation channels grow exponentially, the difference between $\mathcal{O}(MN)$ and $\mathcal{O}(MN^3)$ becomes dramatic, making the optimized approach essential for scalable simulations of large quantum systems.
	
	\subsection{Optimization principles and scalability}
	\label{appxsubsec:Principles}

	The substantial reduction in computational complexity achieved by our approach can be attributed to the inherent sparsity of the jump operators $\hat{A}_k=|j\rangle\langle i|$. This sparsity leads to three crucial simplifications:
	\begin{itemize}
    		\item The term $\hat{A}_k\hat{\rho}\hat{A}_k^\dagger$ simplifies to a single-element update, requiring only $\mathcal{O}(1)$ operations. This is because the jump operator has only one non-zero matrix element, located at the $j$-th row and $i$-th column.
    		\item The product $\hat{A}_k^\dagger\hat{A}_k=|i\rangle\langle i|$ is a diagonal projection operator. Consequently, the anti-commutator $\{\hat{A}_k^\dagger\hat{A}_k,\hat{\rho}\}$ affects only the $i$-th row and the $i$-th column of the density matrix, rather than the entire matrix.
    		\item As a result, the full Lindblad dissipator $\hat{L}_k(\hat{\rho})$ decomposes into a set of localized operations: a single diagonal update (population transfer) combined with independent updates to one specific row and one specific column of the density matrix.
	\end{itemize}

	These simplifications reduce the computational complexity from $\mathcal{O}(N^3)$ per channel in the naive matrix multiplication approach to $\mathcal{O}(N)$ per channel in our optimized implementation. For a system with $M$ dissipation channels, the total complexity scales as $\mathcal{O}(MN)$, compared to $\mathcal{O}(MN^3)$ for the direct approach.

	This linear scaling with respect to the Hilbert space dimension $N$ is particularly significant for quantum simulations, where $N$ grows exponentially with the number of atoms in this paper. The ability to maintain $\mathcal{O}(N)$ scaling per channel enables simulations of open quantum systems with substantially larger Hilbert spaces than would be feasible using conventional matrix multiplication techniques. This optimization strategy has become a cornerstone of efficient numerical methods for the Lindblad master equation in large-scale quantum simulations.
	
	\subsection{Extension to inverse (pumping) processes}
	\label{appxsubsec:Inverse}

	In many physical systems, dissipative channels often coexist with their inverse processes---namely, pumping channels. For example, in quantum optics, spontaneous emission (decay) is often accompanied by pumping processes. In this section, we extend the proposed optimized numerical scheme to handle both dissipative channels and their corresponding pumping channels simultaneously.

	Consider a pair of related channels:
	\begin{itemize}
    		\item Dissipation channel: Jump operator: $\hat{A}_k=|j\rangle\langle i|$. Rate: $\gamma_k$. Direction: $|i\rangle\to|j\rangle$.
	    \item Pumping channel: Jump operator: $\hat{A'}_k=|i\rangle\langle j|$. Rate: $\gamma_k'$ (typically $\gamma_k'<\gamma_k$). Direction: $|j\rangle\to|i\rangle$.
	\end{itemize}
	Both channels share the same pair of states $|i\rangle$ and $|j\rangle$, but with opposite directions and, in general, different rates.

	For each pair of related channels $(\hat{A}_k,\hat{A'}_k)$, the numerical update is implemented as follows. To ensure order independence and strict adherence to the Euler integration scheme, we first make a copy of the density matrix from the previous time step, denoted by $\hat{\rho}_\mathrm{old}=\hat{\rho}$. All operations read from $\hat{\rho}_\mathrm{old}$ and write to $\hat{\rho}$:

	\begin{algorithm}[H]
		\caption{Complete dissipative and pumping channel updates}
		\label{appxalg:ChannelPair}
		\KwIn{Current density matrix $\hat{\rho}$, rates $\gamma_k$ and $\gamma_k'$, time step $\Delta t$, indices $i$, $j$, system size $N$}
		\KwOut{Updated density matrix $\hat{\rho}$}

		\tcc{Make a copy of the current density matrix}
		$\hat{\rho}_\mathrm{old}\leftarrow\hat{\rho}$\;

		\tcc{===== Dissipation channel: $|i\rangle\rightarrow|j\rangle$ with rate $\gamma_k$ =====}
		\tcc{Population transfer from state $i$ to state $j$}
		$\hat{\rho}[j][j]\leftarrow\hat{\rho}[j][j]+\gamma_k\cdot\Delta t\cdot\hat{\rho}_\mathrm{old}[i][i]$\;

		\tcc{Decay of the $i$-th row (using old values)}
		\For{$n\leftarrow 0$ \KwTo $N-1$}{
		    $\hat{\rho}[i][n]\leftarrow\hat{\rho}[i][n]-\left(\frac{\gamma_k\cdot\Delta t}{2}\right)\cdot\hat{\rho}_\mathrm{old}[i][n]$\;
		}

		\tcc{Decay of the $i$-th column (using old values)}
		\For{$m\leftarrow 0$ \KwTo $N-1$}{
		    $\hat{\rho}[m][i]\leftarrow\hat{\rho}[m][i]-\left(\frac{\gamma_k\cdot\Delta t}{2}\right)\cdot\hat{\rho}_\mathrm{old}[m][i]$\;
		}

		\tcc{===== Pumping channel: $|j\rangle\rightarrow|i\rangle$ with rate $\gamma_k'$ =====}
		\tcc{Population transfer from state $j$ to state $i$}
		$\hat{\rho}[i][i]\leftarrow\hat{\rho}[i][i]+\gamma_k'\cdot\Delta t\cdot\hat{\rho}_\mathrm{old}[j][j]$\;

		\tcc{Decay of the $j$-th row (using old values)}
		\For{$n\leftarrow 0$ \KwTo $N-1$}{
		    $\hat{\rho}[j][n]\leftarrow\hat{\rho}[j][n]-\left(\frac{\gamma_k'\cdot\Delta t}{2}\right)\cdot\hat{\rho}_\mathrm{old}[j][n]$\;
		}

		\tcc{Decay of the $j$-th column (using old values)}
		\For{$m\leftarrow 0$ \KwTo $N-1$}{
		    $\hat{\rho}[m][j]\leftarrow\hat{\rho}[m][j]-\left(\frac{\gamma_k'\cdot\Delta t}{2}\right)\cdot\hat{\rho}_\mathrm{old}[m][j]$\;
		}
	\end{algorithm}
	
	Finally, Fig.~\ref{fig:Non-unitary}(b.1)--(b.3) illustrates the effect of the pumping channel on the density matrix.

	Key observations:
	\begin{itemize}
    		\item All operations are based on $\hat{\rho}_\mathrm{old}$, ensuring that the update strictly follows $\hat{\rho}=\hat{\rho}+\Delta t\hat{L}_k(\hat{\rho}_\mathrm{old})$.
	    \item The two channels are independent and can be processed in any order without affecting the final result.
    		\item The net change in the diagonal elements $\hat{\rho}_{i,i}$ and $\hat{\rho}_{j,j}$ is determined by the combined effect of both channels.
	\end{itemize}
	
	For a pair consisting of one dissipative channel and one pumping channel, the operations and their complexities are as follows:
	\begin{table}[h]
		\centering
		\begin{tabular}{|l|c|}
			\hline
			Operation & Complexity \\
			\hline
			Dissipative channel population transfer & $\mathcal{O}(1)$ \\
			Dissipative channel row decay & $\mathcal{O}(N)$ \\
			Dissipative channel column decay & $\mathcal{O}(N)$ \\
			Pumping channel population transfer & $\mathcal{O}(1)$ \\
			Pumping channel row decay & $\mathcal{O}(N)$ \\
			Pumping channel column decay & $\mathcal{O}(N)$ \\
			\hline
			Total per channel pair & $\mathcal{O}(4N+2)=\mathcal{O}(N)$ \\
			\hline
		\end{tabular}
		\caption{{\it Computational complexity breakdown for a pair of dissipative and pumping channels.}}
		\label{appxtab:ComplexityPairBreakdown}
	\end{table}
	
	Despite requiring twice as many operations as a single channel, the asymptotic complexity remains $\mathcal{O}(N)$ per channel pair. For $M$ channel pairs, the total computational complexity scales as $\mathcal{O}(MN)$.
	\begin{table}[h]
		\centering
		\begin{tabular}{|c|c|c|}
			\hline
			Method & Per Channel Pair & $M$ Channel Pairs \\
			\hline
			Direct matrix multiplication & $\mathcal{O}(N^3)$ & $\mathcal{O}(MN^3)$ \\
			\hline
			Optimized (row + column operations) & $\mathcal{O}(N)$ & $\mathcal{O}(MN)$ \\
			\hline
		\end{tabular}
		\caption{{\it Complexity comparison between direct matrix multiplication and the optimized approach for channel pairs.}}
		\label{appxtab:ComplexityPair}
	\end{table}

	The key insight is that the sparsity of the jump operators---each having only one non-zero element---is preserved for both the dissipative and pumping channels. Consequently:
	\begin{itemize}
    		\item $\hat{A}_k\hat{\rho}\hat{A}_k^\dagger$ and $\hat{A'}_k\hat{\rho}\hat{A'}_k^\dagger$ both reduce to single-element updates, each requiring only $\mathcal{O}(1)$ operations.
	    \item $\hat{A}_k^\dagger\hat{A}_k=|i\rangle\langle i|$ and $\hat{A'}_k^\dagger\hat{A'}_k=|j\rangle\langle j|$ are both diagonal projectors. Consequently, the anti-commutator terms affect only a single row and a single column, each requiring $\mathcal{O}(N)$ operations.
	\end{itemize}

	Therefore, even when considering pairs of channels, the total complexity scales as $\mathcal{O}(MN)$ for $M$ channel pairs, compared to $\mathcal{O}(MN^3)$ for the naive matrix multiplication approach. This linear scaling in $N$ is crucial for simulating systems with large Hilbert space dimensions.
	
	\subsection{Parallelization of non-unitary terms}
	\label{appxsubsec:Parallelization}
	
	Following the optimizations for non-unitary terms described above, the numerical update of a single dissipative channel can be decomposed into three basic operations: a point operation, a row operation, and a column operation. The point operation implements the population transfer from state $|i\rangle$ to state $|j\rangle$, while the row and column operations correspond to the dissipative effects arising from the anti-commutator $-\dfrac{1}{2}\{\hat{\rho}, \hat{A}_k^\dagger\hat{A}_k\}$.

	From the perspective of parallel computing, these three operations have distinctly different communication characteristics:
	\begin{itemize}
		\item Point operation: This operation updates the matrix element $\hat{\rho}_{j,j}$, requiring reading the current value of $\hat{\rho}_{i,i}$. In a distributed storage environment, if $\hat{\rho}_{i,i}$ and $\hat{\rho}_{j,j}$ are stored on different processors, a data transfer is required.
		\item Row operation: This operation decays all elements in the $i$-th row of the density matrix. Although the row elements may be distributed across multiple processors due to the block-based matrix partitioning, each processor can independently update its locally stored row segment without cross-processor communication.
		\item Column operation: This operation decays all elements in the $i$-th column of the density matrix. Similarly, each processor can independently update its local column segment without communication.
	\end{itemize}
	
	\begin{figure}
		\centering
		\includegraphics[width=1.\textwidth]{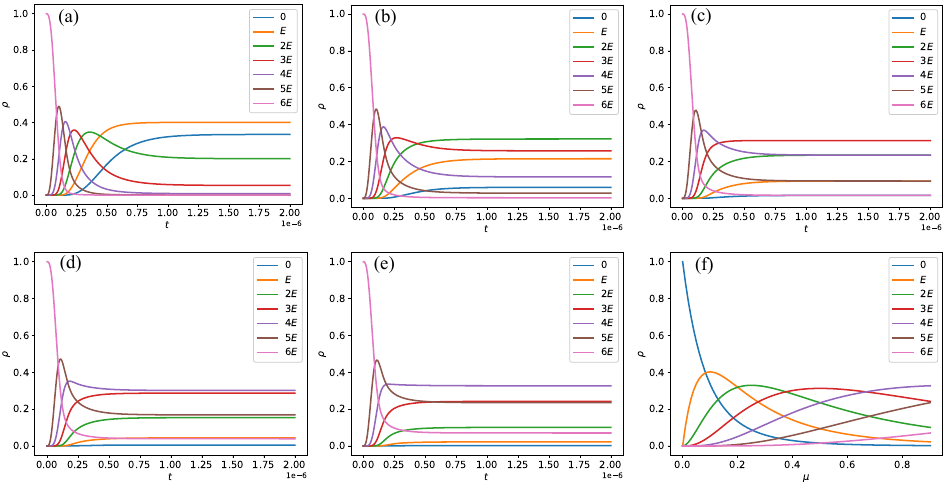} 
		\caption{(online color) {\it Dissipative-pumped dynamics of the TCM with $N_\mathrm{atoms}=6$ two-level atoms.} Same format as Fig.~\ref{fig:Dissipative+Pumped-5Atoms} for $N_\mathrm{atoms}=6$. Panels (a)--(e): Time evolution for $\mu=0.1, 0.3, 0.5, 0.7, 0.9$. Panel (f): Steady-state population distribution versus $\mu$ for energy levels $0$ to $6E$.}
		\label{appxfig:Dissipative+Pumped-6Atoms}
	\end{figure}
	
	\begin{figure}
		\centering
		\includegraphics[width=1.\textwidth]{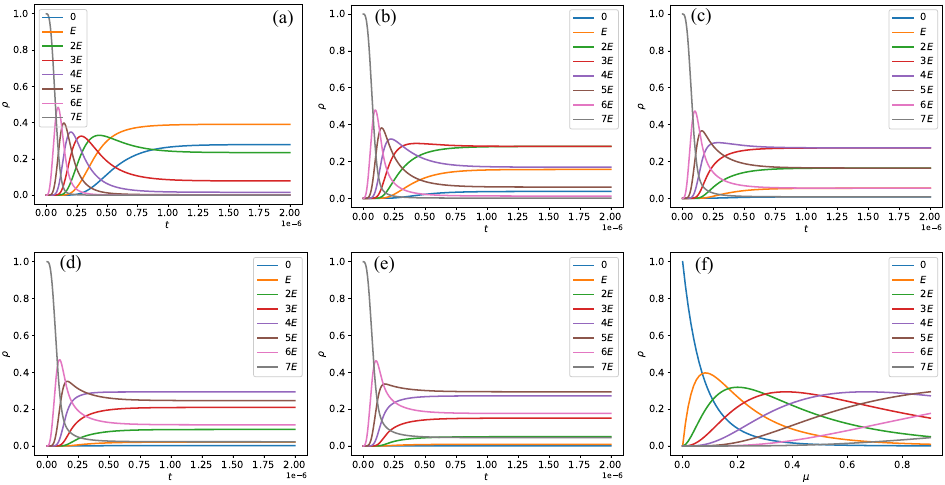} 
		\caption{(online color) {\it Dissipative-pumped dynamics of the TCM with $N_\mathrm{atoms}=7$ two-level atoms.} Same format as Fig.~\ref{fig:Dissipative+Pumped-5Atoms} for $N_\mathrm{atoms}=7$. Panels (a)--(e): Time evolution for $\mu=0.1, 0.3, 0.5, 0.7, 0.9$. Panel (f): Steady-state population distribution versus $\mu$ for energy levels $0$ to $7E$.}
		\label{appxfig:Dissipative+Pumped-7Atoms}
	\end{figure}
	
	\begin{figure}
		\centering
		\includegraphics[width=1.\textwidth]{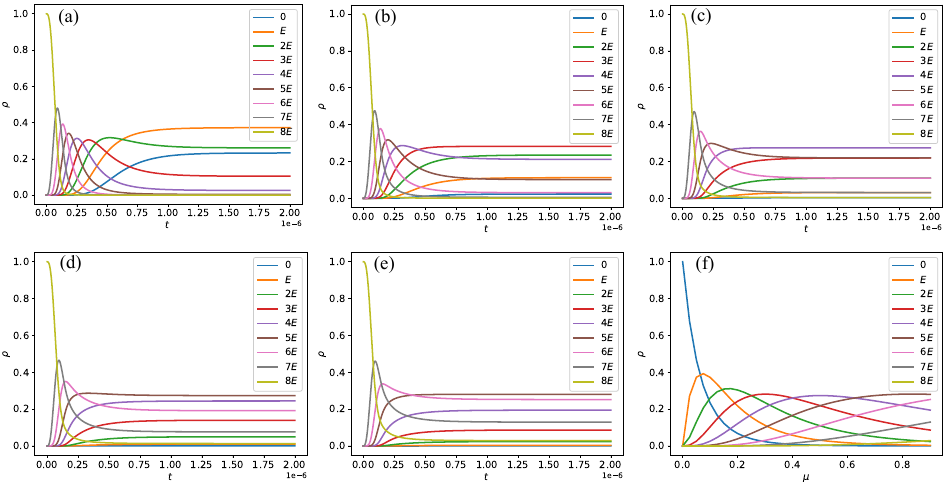} 
		\caption{(online color) {\it Dissipative-pumped dynamics of the TCM with $N_\mathrm{atoms}=8$ two-level atoms.} Same format as Fig.~\ref{fig:Dissipative+Pumped-5Atoms} for $N_\mathrm{atoms}=8$. Panels (a)--(e): Time evolution for $\mu=0.1, 0.3, 0.5, 0.7, 0.9$. Panel (f): Steady-state population distribution versus $\mu$ for energy levels $0$ to $8E$.}
		\label{appxfig:Dissipative+Pumped-8Atoms}
	\end{figure}
	
	\begin{figure}
		\centering
		\includegraphics[width=1.\textwidth]{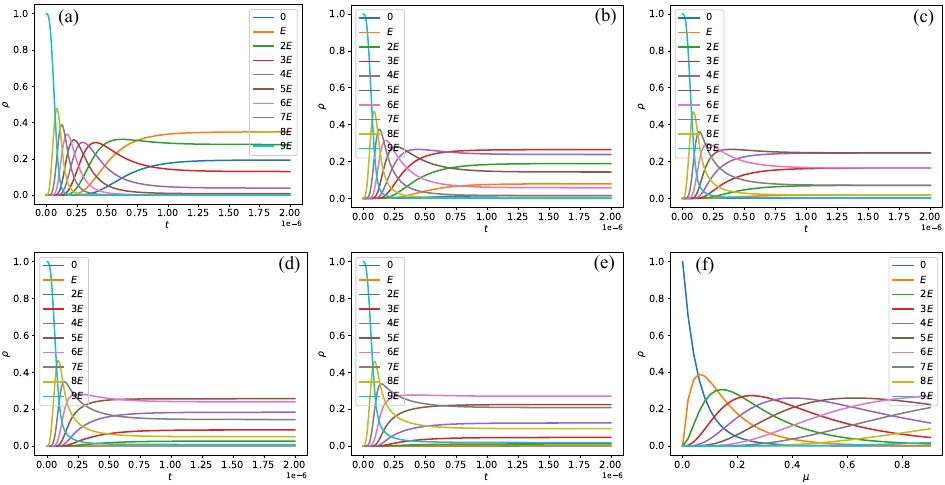} 
		\caption{(online color) {\it Dissipative-pumped dynamics of the TCM with $N_\mathrm{atoms}=9$ two-level atoms.} Same format as Fig.~\ref{fig:Dissipative+Pumped-5Atoms} for $N_\mathrm{atoms}=9$. Panels (a)--(e): Time evolution for $\mu=0.1, 0.3, 0.5, 0.7, 0.9$. Panel (f): Steady-state population distribution versus $\mu$ for energy levels $0$ to $9E$.}
		\label{appxfig:Dissipative+Pumped-9Atoms}
	\end{figure}

	Therefore, in the entire non-unitary term update process, only point operations may trigger data exchange between processors, while row and column operations can be completed locally. This property has significant implications for parallelization:
	\begin{itemize}
		\item Minimized communication overhead: At most one communication (the point operation) is required per time step per channel, and the communication volume is independent of the matrix dimension $N$, scaling as $\mathcal{O}(1)$.
		\item Balanced computational load: Row and column operations naturally decompose into parallel processing of local data blocks by each processor, with the computational load scaling linearly with the number of processors.
		\item Reduced parallelization complexity: No complex communication topology or global synchronization strategy is needed; standard block distribution combined with simple point-to-point communication suffices to achieve efficient parallelism.
	\end{itemize}

	In summary, the optimized decomposition of non-unitary terms not only reduces the computational complexity of a single channel from $\mathcal{O}(N^3)$ to $\mathcal{O}(N)$ but, more importantly, restricts communication requirements to the single-element level, ensuring good scalability on distributed-memory architectures. For a system with $M$ dissipative channels, the total communication volume scales as $\mathcal{O}(M)$, and the computational complexity scales as $\mathcal{O}(MN)$, laying the foundation for parallel simulations of large-scale open quantum systems.
	
	\begin{figure}
		\centering
        \includegraphics[width=1.\textwidth]{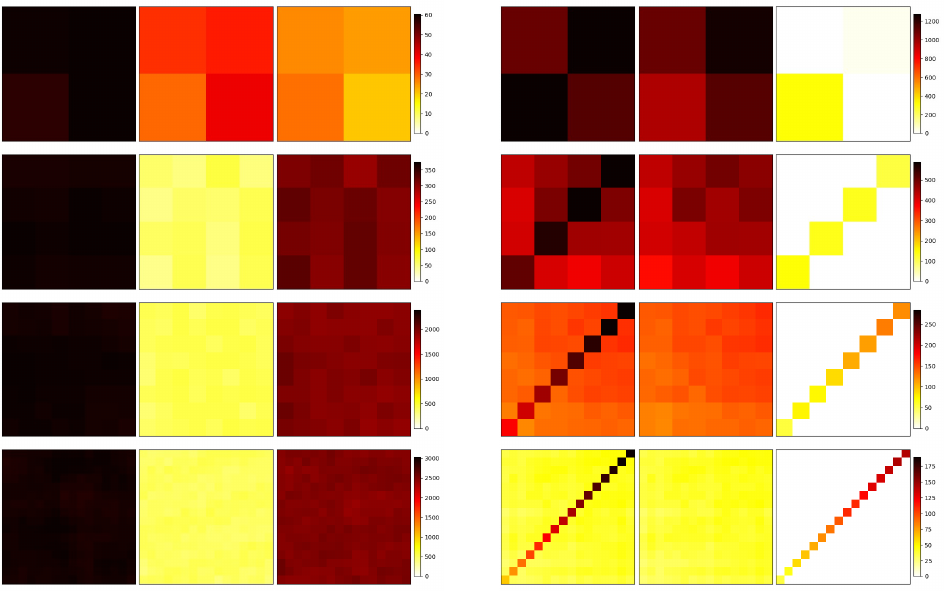}
        \caption{(online color) {\it The case of 6 atoms.} The layout follows the same structure as Fig.~\ref{fig:Heatmap}.}
        \label{appxfig:6Atoms}
	\end{figure}
	
	\section{Supplementary results for dissipative-pumped dynamics}
	\label{appxsec:DissPumpDynamics}

	This appendix provides the dissipative-pumped dynamics for intermediate system sizes ($N_{\mathrm{atoms}} = 6, 7, 8, 9$), which follow the same format as Fig.~\ref{fig:Dissipative+Pumped-5Atoms} in the main text. These results are included for completeness; the main text focuses on the smallest ($N_{\mathrm{atoms}}=5$) and largest ($N_{\mathrm{atoms}}=10$) systems to illustrate the size-dependent trends.

	These figures (Figs.~\ref{fig:Dissipative+Pumped-5Atoms}--\ref{fig:Dissipative+Pumped-10Atoms} and Figs.~\ref{appxfig:Dissipative+Pumped-6Atoms}--\ref{appxfig:Dissipative+Pumped-9Atoms}) demonstrate that the cascade behavior and the emergence of a non-equilibrium steady state persist across all system sizes from $N_{\mathrm{atoms}}=5$ to $10$. As $N_{\mathrm{atoms}}$ increases, the following trends are observed:
	\begin{itemize}
		\item The population cascade extends to higher energy levels, with the maximum populated energy level increasing with $N_{\mathrm{atoms}}$.
		\item The characteristic $\mu$ value at which each energy level reaches its peak shifts systematically with $N_{\mathrm{atoms}}$.
	\end{itemize}
	
	\section{Breakdown of the computation time for unitary and non-unitary terms across different processor grid sizes for systems with 6 to 10 atoms}
	\label{appxsec:Breakdown}
	
	\begin{figure}
		\centering
        \includegraphics[width=1.\textwidth]{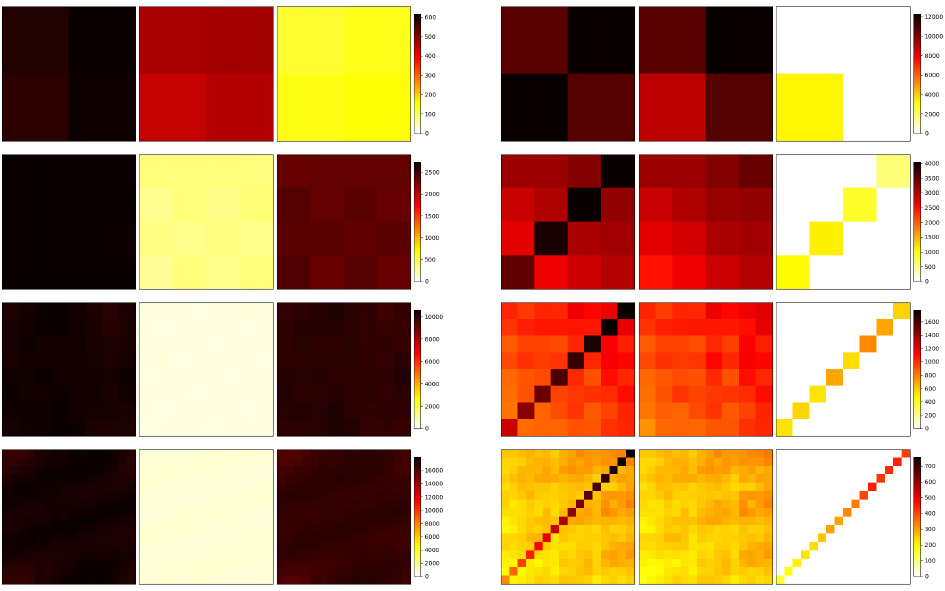}
        \caption{(online color) {\it The case of 7 atoms.} The layout follows the same structure as Fig.~\ref{fig:Heatmap}.}
        \label{appxfig:7Atoms}
	\end{figure}
	
	For the 6-atom system, the reduced Hamiltonian dimension is 729, and the total number of dissipation channels is 1458. As shown in Fig.~\ref{appxfig:6Atoms}, as the processor grid size increases from $2\times 2$ to $16\times 16$, the time required to solve the unitary term over 10 iterations increases from 60 seconds to 374 seconds, then to 2387 seconds, and finally to 3030 seconds. The overall trend in time cost shows a significant increase. This is mainly due to the sharp rise in communication overhead caused by the increasing number of cross-processor communications. Conversely, the time required to solve the non-unitary term over 10 iterations decreases rapidly: the time required to solve the non-unitary term over 10 iterations decreases rapidly: the maximum value of the color bars drops from 1275 seconds to 588 seconds, then to 284 seconds, and finally to 189 seconds. This is because the non-unitary term involves minimal cross-processor communication. Moreover, as the processor grid size increases, the number of multiply-accumulate operations assigned to each processor decreases, leading to significant time savings.
	
	For the 7-atom system, the reduced Hamiltonian dimension is 2187, and the total number of dissipation channels is 5103. As shown in Fig.~\ref{appxfig:7Atoms}, as the processor grid size increases from $2\times 2$ to $16\times 16$, the time required to solve the unitary term over 10 iterations increases from 614 seconds to 2729 seconds, then to 10583 seconds, and finally to 17968 seconds. The overall trend in time cost shows a significant increase. The time required to solve the non-unitary term over 10 iterations decreases rapidly: the maximum value of the color bars drops from 12255 seconds to 4038 seconds, then to 1772 seconds, and finally to 757 seconds.
	
\begin{figure}
		\centering
        \includegraphics[width=1.\textwidth]{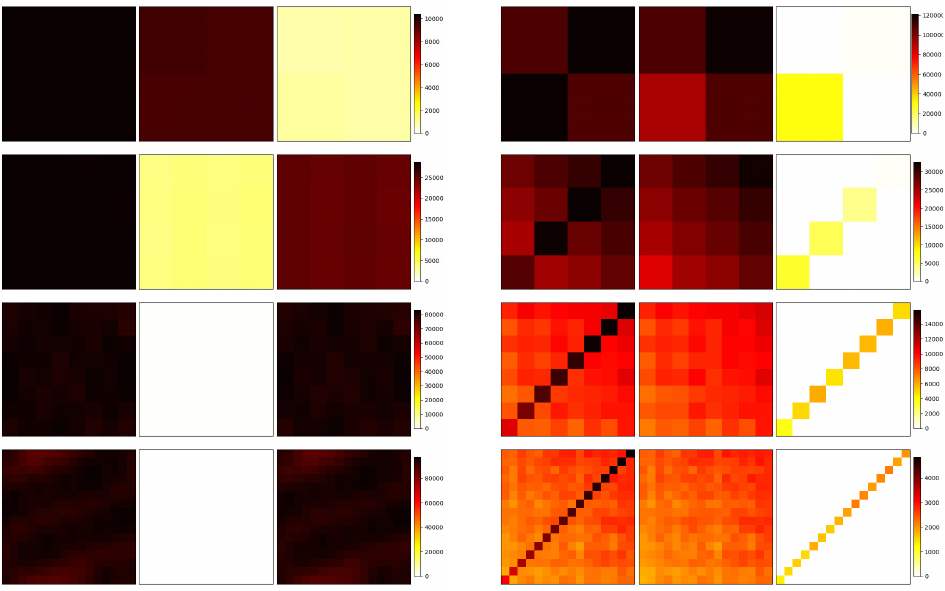}
        \caption{(online color) {\it The case of 8 atoms.} The layout follows the same structure as Fig.~\ref{fig:Heatmap}.}
        \label{appxfig:8Atoms}
	\end{figure}

	For the 8-atom system, the reduced Hamiltonian dimension is 6561, and the total number of dissipation channels is 17496. As shown in Fig.~\ref{appxfig:8Atoms}, as the processor grid size increases from $2\times 2$ to $16\times 16$, the time required to solve the unitary term over 10 iterations increases from 10377 seconds to 28710 seconds, then to 83126 seconds, and finally to 97335 seconds. The overall trend in time cost shows a significant increase. The time required to solve the non-unitary term over 10 iterations decreases rapidly: the maximum value of the color bars drops from 120973 seconds to 32603 seconds, then to 15868 seconds, and finally to 4842 seconds.
	
	\begin{figure}
		\centering
        \includegraphics[width=1.\textwidth]{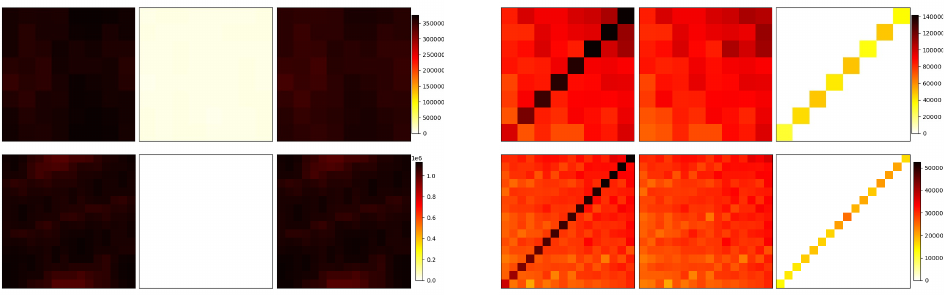}
        \caption{(online color) {\it The case of 9 atoms.} Each row corresponds to a different processor grid configuration: from top to bottom, $8\times 8$ and $16\times 16$ processors. The rest of the layout follows the same structure as in Fig.~\ref{fig:Heatmap}.}
        \label{appxfig:9Atoms}
	\end{figure}
	
	For the 9-atom system, the reduced Hamiltonian dimension is 19683, and the total number of dissipation channels is 59049. As shown in Fig.~\ref{appxfig:9Atoms}, as the processor grid size increases from $8\times 8$ to $16\times 16$, the time required to solve the unitary term over 10 iterations increases from $3.76\times 10^5$ seconds to $1.13\times 10^6$ seconds. The overall trend in time cost shows no significant increase. The time required to solve the non-unitary term over 10 iterations decreases: the maximum value of the color bars drops from $1.42\times 10^5$ seconds to $5.26\times 10^4$ seconds.
	
\begin{figure}
		\centering
        \includegraphics[width=1.\textwidth]{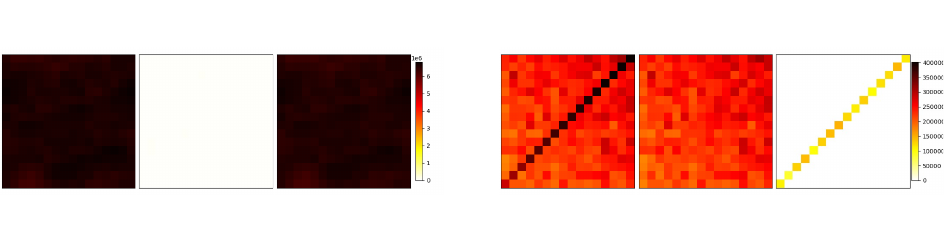}
        \caption{(online color) {\it The case of 10 atoms.} A single row corresponds to a grid configuration with $16\times 16$ processors. The rest of the layout follows the same structure as Fig.~\ref{fig:Heatmap}.}
        \label{appxfig:10Atoms}
	\end{figure}

	For the 10-atom system, the reduced Hamiltonian dimension is 59049, and the total number of dissipation channels is 196830. As shown in Fig.~\ref{appxfig:10Atoms}, for the grid configuration with $16\times 16$ processors, the time required to solve the unitary term over 10 iterations equals $6.82\times 10^6$ seconds. The time required to solve the non-unitary term over 10 iterations decreases to $4.02\times 10^5$ seconds.
	
	\bibliography{bibliography}

@article{McArdle2020,
  title = {Quantum computational chemistry},
  author = {McArdle, Sam and Endo, Suguru and Aspuru-Guzik, Al\'an and Benjamin, Simon C. and Yuan, Xiao},
  journal = {Rev. Mod. Phys.},
  volume = {92},
  issue = {1},
  pages = {015003},
  numpages = {51},
  year = {2020},
  month = {Mar},
  publisher = {American Physical Society},
  doi = {10.1103/RevModPhys.92.015003},
  url = {https://link.aps.org/doi/10.1103/RevModPhys.92.015003}
}

@article{Baiardi2023,
author = {Baiardi, Alberto and Christandl, Matthias and Reiher, Markus},
title = {Quantum Computing for Molecular Biology},
journal = {ChemBioChem},
volume = {24},
number = {13},
pages = {e202300120},
doi = {https://doi.org/10.1002/cbic.202300120},
url = {https://chemistry-europe.onlinelibrary.wiley.com/doi/abs/10.1002/cbic.202300120},
year = {2023}
}

@book{Albuquerque2021,
title = {Quantum Chemistry Simulation of Biological Molecules},
publisher = {Cambridge University Press},
year = {2021},
author = {Albuquerque, EL and Fulco, UL and Caetano, EWS and Freire, VN},
}

@book{Bellman1957,
title = {Dynamic Programming},
publisher = {Princeton University Press},
year = {1957},
author = {Bellman, R. E.},
}

@book{Bellman1961,
title = {Adaptive control processes: a guided tour},
publisher = {Princeton University Press},
year = {1961},
author = {Bellman, R. E.},
}

@article{Rabi1936,
  title = {On the Process of Space Quantization},
  author = {Rabi, I. I.},
  journal = {Phys. Rev.},
  volume = {49},
  issue = {4},
  pages = {324--328},
  numpages = {0},
  year = {1936},
  month = {Feb},
  publisher = {American Physical Society},
  doi = {10.1103/PhysRev.49.324},
  url = {https://link.aps.org/doi/10.1103/PhysRev.49.324}
}

@article{Rabi1937,
  title = {Space Quantization in a Gyrating Magnetic Field},
  author = {Rabi, I. I.},
  journal = {Phys. Rev.},
  volume = {51},
  issue = {8},
  pages = {652--654},
  numpages = {0},
  year = {1937},
  month = {Apr},
  publisher = {American Physical Society},
  doi = {10.1103/PhysRev.51.652},
  url = {https://link.aps.org/doi/10.1103/PhysRev.51.652}
}

@article{Dicke1954,
  title = {Coherence in Spontaneous Radiation Processes},
  author = {Dicke, R. H.},
  journal = {Phys. Rev.},
  volume = {93},
  issue = {1},
  pages = {99--110},
  numpages = {0},
  year = {1954},
  month = {Jan},
  publisher = {American Physical Society},
  doi = {10.1103/PhysRev.93.99},
  url = {https://link.aps.org/doi/10.1103/PhysRev.93.99}
}

@article{Hopfield1958,
  title = {Theory of the Contribution of Excitons to the Complex Dielectric Constant of Crystals},
  author = {Hopfield, J. J.},
  journal = {Phys. Rev.},
  volume = {112},
  issue = {5},
  pages = {1555--1567},
  numpages = {0},
  year = {1958},
  month = {Dec},
  publisher = {American Physical Society},
  doi = {10.1103/PhysRev.112.1555},
  url = {https://link.aps.org/doi/10.1103/PhysRev.112.1555}
}

@article{Casanova2010,
  title = {Deep Strong Coupling Regime of the {J}aynes--{C}ummings Model},
  author = {Casanova, J. and Romero, G. and Lizuain, I. and Garc\'{\i}a-Ripoll, J. J. and Solano, E.},
  journal = {Phys. Rev. Lett.},
  volume = {105},
  issue = {26},
  pages = {263603},
  numpages = {4},
  year = {2010},
  month = {Dec},
  publisher = {American Physical Society},
  doi = {10.1103/PhysRevLett.105.263603},
  url = {https://link.aps.org/doi/10.1103/PhysRevLett.105.263603}
}

@article{Haroche2013,
  title = {Nobel Lecture: Controlling photons in a box and exploring the quantum to classical boundary},
  author = {Haroche, Serge},
  journal = {Rev. Mod. Phys.},
  volume = {85},
  issue = {3},
  pages = {1083--1102},
  numpages = {0},
  year = {2013},
  month = {Jul},
  publisher = {American Physical Society},
  doi = {10.1103/RevModPhys.85.1083},
  url = {https://link.aps.org/doi/10.1103/RevModPhys.85.1083}
}

@article{Gu2017,
  title = {Microwave photonics with superconducting quantum circuits},
  journal = {Physics Reports},
  volume = {718-719},
  pages = {1-102},
  year = {2017},
  issn = {0370-1573},
  doi = {https://doi.org/10.1016/j.physrep.2017.10.002},
  url = {https://www.sciencedirect.com/science/article/pii/S0370157317303290},
  author = {Xiu Gu and Anton Frisk Kockum and Adam Miranowicz and Yu-xi Liu and Franco Nori},
}

@Inbook{Kockum2019,
  author="Kockum, Anton Frisk and Nori, Franco",
  editor="Tafuri, Francesco",
  title="Quantum Bits with {J}osephson Junctions",
  bookTitle="Fundamentals and Frontiers of the Josephson Effect",
  year="2019",
  publisher="Springer International Publishing",
  address="Cham",
  pages="703--741",
  isbn="978-3-030-20726-7",
  doi="10.1007/978-3-030-20726-7_17",
  url="https://doi.org/10.1007/978-3-030-20726-7_17"
}

@article{KockumMiranowicz2019,
  title = {Ultrastrong coupling between light and matter},
  journal = {Nature Reviews Physics},
  volume = {1},
  pages = {19-40},
  year = {2019},
  doi = {10.1038/s42254-018-0006-2},
  url = {https://doi.org/10.1038/s42254-018-0006-20},
  author = {Frisk Kockum, Anton and Miranowicz, Adam and De Liberato, Simone and Savasta, Salvatore and Nori, Franco},
}

@article{Forn-Diaz2019,
  title = {Ultrastrong coupling regimes of light-matter interaction},
  author = {Forn-D\'{\i}az, P. and Lamata, L. and Rico, E. and Kono, J. and Solano, E.},
  journal = {Rev. Mod. Phys.},
  volume = {91},
  issue = {2},
  pages = {025005},
  numpages = {48},
  year = {2019},
  month = {Jun},
  publisher = {American Physical Society},
  doi = {10.1103/RevModPhys.91.025005},
  url = {https://link.aps.org/doi/10.1103/RevModPhys.91.025005}
}

@ARTICLE{Jaynes1963,
  author={Jaynes, E. T. and Cummings, F. W.},
  journal={Proceedings of the IEEE}, 
  title={Comparison of quantum and semiclassical radiation theories with application to the beam maser}, 
  year={1963},
  volume={51},
  number={1},
  pages={89-109},
  doi={10.1109/PROC.1963.1664},
  url={https://doi.org/10.1109/PROC.1963.1664}
}

@article{Tavis1968,
  title = {Exact Solution for an $N$-Molecule---Radiation-Field {H}amiltonian},
  author = {Tavis, Michael and Cummings, Frederick W.},
  journal = {Phys. Rev.},
  volume = {170},
  issue = {2},
  pages = {379--384},
  numpages = {0},
  year = {1968},
  month = {Jun},
  publisher = {American Physical Society},
  doi = {10.1103/PhysRev.170.379},
  url = {https://link.aps.org/doi/10.1103/PhysRev.170.379}
}

@article{Angelakis2007,
  title = {Photon-blockade-induced Mott transitions and {$XY$} spin models in coupled cavity arrays},
  author = {Angelakis, Dimitris G. and Santos, Marcelo Franca and Bose, Sougato},
  journal = {Phys. Rev. A},
  volume = {76},
  issue = {3},
  pages = {031805},
  numpages = {4},
  year = {2007},
  month = {Sep},
  publisher = {American Physical Society},
  doi = {10.1103/PhysRevA.76.031805},
  url = {https://link.aps.org/doi/10.1103/PhysRevA.76.031805}
}

@article{Prasad2018,
  title = {Effective Three-Body Interactions in {J}aynes-{C}ummings-{H}ubbard Systems},
  author = {Prasad, S.B. and Martin, A.M.},
  journal = {Sci Rep},
  volume = {8},
  issue = {1},
  pages = {16253},
  year = {2018},
  month = {Nov},
  doi = {10.1038/s41598-018-33907-9},
  url = {https://doi.org/10.1038/s41598-018-33907-9}
}

@article{Wei2021,
  title = {Quantum Monte Carlo study of superradiant supersolid of light in the extended {J}aynes--{C}ummings--{H}ubbard model},
  author = {Wei, Huanhuan and Zhang, Jie and Greschner, Sebastian and Scott, Tony C and Zhang, Wanzhou},
  journal = {Phys. Rev. B},
  volume = {103},
  issue = {18},
  pages = {184501},
  numpages = {11},
  year = {2021},
  month = {May},
  publisher = {American Physical Society},
  doi = {10.1103/PhysRevB.103.184501},
  url = {https://link.aps.org/doi/10.1103/PhysRevB.103.184501}
}

@article{OzhigovYI2020,
doi = {10.1070/QEL17320},
url = {https://doi.org/10.1070/QEL17320},
year = {2020},
month = {oct},
publisher = {Kvantovaya Elektronika, Turpion Ltd and IOP Publishing},
volume = {50},
number = {10},
pages = {947},
author = {Ozhigov, Yu.I.},
title = {Quantum gates on asynchronous atomic excitations},
journal = {Quantum Electronics}
}

@article{Smith2021,
  title = {Exact $k$-body representation of the {J}aynes--{C}ummings interaction in the dressed basis: Insight into many-body phenomena with light},
  author = {Smith, Kevin C. and Bhattacharya, Aniruddha and Masiello, David J.},
  journal = {Phys. Rev. A},
  volume = {104},
  issue = {1},
  pages = {013707},
  numpages = {23},
  year = {2021},
  month = {Jul},
  publisher = {American Physical Society},
  doi = {10.1103/PhysRevA.104.013707},
  url = {https://link.aps.org/doi/10.1103/PhysRevA.104.013707}
}

@article{Dull2021,
  title = {Quality of Control in the {T}avis--{C}ummings--{H}ubbard Model},
  author = {Düll, R. and Kulagin, A. and Lee, L. and Ozhigov, Y. and Miao, H. and Zheng, K.},
  journal = {Computational Mathematics and Modeling},
  volume = {32},
  issue = {1},
  pages = {75-85},
  year = {2021},
  doi = {10.1007/s10598-021-09517-y},
  url = {https://doi.org/10.1007/s10598-021-09517-y}
}

@article{Miao2024,
author = {Miao, Hui-hui},
title = {Investigating entropic dynamics of multiqubit cavity {QED} system},
journal = {Advanced Quantum Technologies},
volume = {7},
issue = {12},
pages = {2400246},
year = {2024},
doi = {10.1002/qute.202400246},
URL = {http://dx.doi.org/10.1002/qute.202400246},
}

@article{MiaoLi2025,
author = {Miao, Hui-hui and Li, Wanshun},
title = {Entanglement and quantum discord in the cavity {QED} models},
journal = {Heliyon},
volume = {11},
issue = {1},
pages = {e41194},
year = {2025},
doi = {10.1016/j.heliyon.2024.e41194},
URL = {https://doi.org/10.1016/j.heliyon.2024.e41194},
}

@article{You2024,
title = {Description of the non-Markovian dynamics of atoms in terms of a pure state},
journal = {Comput Math Model},
volume = {34},
pages = {75-84},
year = {2024},
doi = {https://doi.org/10.1007/s10598-024-09596-7},
author = {Ozhigov, Yuri Igorevich and You, Jiangchuan}
}

@article{MiaoOzhigov2024,
author = {Miao, Hui-hui and Ozhigov, Yuri Igorevich},
title = {Distributed computing quantum unitary evolution},
journal = {Lobachevskii Journal of Mathematics},
volume = {45},
issue = {7},
pages = {3121-3129},
year = {2024},
doi = {10.1134/S1995080224603904},
URL = {http://dx.doi.org/10.1134/S1995080224603904},
}

@article{LiMiao2024,
author = {Li, Wanshun and Miao, Hui-hui and Ozhigov, Yuri Igorevich},
title = {Supercomputer model of finite-dimensional quantum electrodynamics applications},
journal = {Lobachevskii Journal of Mathematics},
volume = {45},
issue = {7},
pages = {3097-3106},
year = {2024},
doi = {10.1134/S1995080224603849},
URL = {http://dx.doi.org/10.1134/S1995080224603849},
}

@Article{Fink2008,
AUTHOR = {Fink, J. M. and Göppl, M. and Baur, M. and Bianchetti, R. and Leek, P. J. and Blais, A. and Wallraff, A.},
TITLE = {Climbing the {J}aynes--{C}ummings ladder and observing its nonlinearity in a cavity {QED} system},
JOURNAL = {Nature},
VOLUME = {454},
YEAR = {2008},
PAGES = {315--318},
URL = {https://doi.org/10.1038/nature07112},
DOI = {10.1038/nature07112},
}

@Article{Dombi2015,
AUTHOR = {Dombi, A. and Vukics, A. and Domokos, P.},
TITLE = {Bistability effect in the extreme strong coupling regime of the {J}aynes-{C}ummings model},
JOURNAL = {Eur. Phys. J.},
VOLUME = {69},
YEAR = {2015},
number = {60},
DOI = {10.1140/epjd/e2015-50861-9},
URL = {https://doi.org/10.1140/epjd/e2015-50861-9},
}

@book{Larson2021,
author = {Larson, Jonas and Mavrogordatos, Themistoklis},
title = {The {J}aynes--{C}ummings Model and Its Descendants},
publisher = {IOP Publishing},
year = {2021},
series = {2053-2563},
isbn = {978-0-7503-3447-1},
url = {https://doi.org/10.1088/978-0-7503-3447-1},
doi = {10.1088/978-0-7503-3447-1}
}

@article{Kirton2019,
author = {Kirton, Peter and Roses, Mor M. and Keeling, Jonathan and Dalla Torre, Emanuele G.},
title = {Introduction to the Dicke Model: From Equilibrium to Nonequilibrium, and Vice Versa},
journal = {Advanced Quantum Technologies},
volume = {2},
number = {1-2},
pages = {1800043},
doi = {https://doi.org/10.1002/qute.201800043},
url = {https://advanced.onlinelibrary.wiley.com/doi/abs/10.1002/qute.201800043},
year = {2019}
}

@article{Wu2007,
  title = {Strong-Coupling Theory of Periodically Driven Two-Level Systems},
  author = {Wu, Ying and Yang, Xiaoxue},
  journal = {Phys. Rev. Lett.},
  volume = {98},
  issue = {1},
  pages = {013601},
  numpages = {4},
  year = {2007},
  month = {Jan},
  publisher = {American Physical Society},
  doi = {10.1103/PhysRevLett.98.013601},
  url = {https://link.aps.org/doi/10.1103/PhysRevLett.98.013601}
}

@article{Miao2023,
title = {Using a modified version of the Tavis-Cummings-Hubbard model to simulate the formation of neutral hydrogen molecule},
journal = {Physica A: Statistical Mechanics and its Applications},
pages = {128851},
year = {2023},
issn = {0378-4371},
doi = {https://doi.org/10.1016/j.physa.2023.128851},
url = {https://www.sciencedirect.com/science/article/pii/S0378437123004065},
author = {Hui-hui Miao and Yuri Igorevich Ozhigov}
}

@book{Breuer2002,
  author="Breuer, Heinz-Peter and Petruccione, Francesco and others",
  bookTitle="The theory of open quantum systems",
  year="2002",
  publisher="Oxford University Press",
  url="http://refhub.elsevier.com/S0378-4371(22)00557-X/sb42"
}

@article{Alicki1979,
	doi = {10.1088/0305-4470/12/5/007},
	url = {https://doi.org/10.1088/0305-4470/12/5/007},
	year = 1979,
	month = {may},
	publisher = {{IOP} Publishing},
	volume = {12},
	number = {5},
	pages = {L103--L107},
	author = {R Alicki},
	title = {The quantum open system as a model of the heat engine},
	journal = {Journal of Physics A: Mathematical and General}
}

@Article{Kosloff2013,
AUTHOR = {Kosloff, Ronnie},
TITLE = {Quantum Thermodynamics: A Dynamical Viewpoint},
JOURNAL = {Entropy},
VOLUME = {15},
YEAR = {2013},
NUMBER = {6},
PAGES = {2100--2128},
URL = {https://www.mdpi.com/1099-4300/15/6/2100},
ISSN = {1099-4300},
}

@phdthesis{cannon1969,
title = {{PhD} Dissertation: A cellular computer to implement the kalman filter algorithm},
year = {1969},
author = {Cannon, L. E.},
}

@article{Moler2003,
author = {Moler, Cleve and Van Loan, Charles},
title = {Nineteen Dubious Ways to Compute the Exponential of a Matrix, Twenty-Five Years Later},
journal = {SIAM Review},
volume = {45},
number = {1},
pages = {3-49},
year = {2003},
doi = {10.1137/S00361445024180},
URL = {https://doi.org/10.1137/S00361445024180},
eprint = {https://doi.org/10.1137/S00361445024180}
}

@article{Sidje1998,
author = {Sidje, Roger B.},
title = {Expokit: a software package for computing matrix exponentials},
year = {1998},
issue_date = {March 1998},
publisher = {Association for Computing Machinery},
address = {New York, NY, USA},
volume = {24},
number = {1},
issn = {0098-3500},
url = {https://doi.org/10.1145/285861.285868},
doi = {10.1145/285861.285868},
journal = {ACM Trans. Math. Softw.},
month = {mar},
pages = {130–156},
numpages = {27}
}

\end{document}